\documentclass[twocolumn,showpacs,
  nofootinbib,aps,superscriptaddress,
  eqsecnum,prd,notitlepage,showkeys,10pt]{revtex4-1}

\usepackage{amssymb}
\usepackage{amsmath}
\usepackage{graphicx}
\graphicspath{ {./images/} }
\usepackage{dcolumn}
\usepackage{natbib}

\newtheorem{theorem}{Theorem}[section]

\usepackage[colorlinks]{hyperref}
\makeatletter
\newcommand{\setword}[2]{%
  \phantomsection
  #1\def\@currentlabel{\unexpanded{#1}}\label{#2}%
}
\makeatother

\newcommand{\bra}[1]{\langle#1\rvert} 
\newcommand{\ket}[1]{\lvert#1\rangle} 

\begin{document}

\title{A Survey of Methods for Mitigating Barren Plateaus for Parameterized Quantum Circuits}
\author{Michelle Gelman}
\affiliation{Department of Computer Science, The University of Texas at Austin}

\begin{abstract}
Barren Plateaus are a formidable challenge for hybrid quantum-classical algorithms that lead to flat plateaus in the loss function landscape making it difficult to take advantage of the expressive power of parameterized quantum circuits with gradient-based methods. Like in classical neural network models, parameterized quantum circuits suffer the same vanishing gradient issue due to large parameter spaces with non-convex landscapes. In this review, we present an overview of the different genesis for barren plateaus, mathematical formalisms of common themes around barren plateaus, and dives into gradients. The central objective is to provide a conceptual perspective between classical and quantum interpretations of vanishing gradients as well as dive into techniques involving cost functions, entanglement, and initialization strategies to mitigate barren plateaus. Addressing barren plateaus paves the way towards feasibility of many classically intractable applications for quantum simulation, optimization, chemistry, and quantum machine learning. 
\end{abstract}

\maketitle

\section{Introduction}
\indent
Variational quantum algorithms (VQAs) are an emerging hybrid classic-quantum technique in which a trial wave function is constructed as a function of parameterized unitary gates to minimize some objective function via an observable. These algorithms use quantum subroutines where measurement statistics are evaluated with classical processing to update the circuit parameters of the trial wave function with an iterative optimization loop~~\cite{du_quantum_2022}. In all, a VQA consists of an initial state and parameterized unitaries, $U(\theta)$, that model circuit structure, a problem-specific cost function that encodes the optimization problem, and training procedure for parameter updating. These optimization algorithms are akin to finding the lowest expectation energy ground state which serves as a model for many domains of interest including quantum simulation, condensed matter physics, quantum chemistry, combinatorial optimization, and machine learning~\cite{leone_practical_2022, cao_quantum_2019, noauthor_practical_nodate}. As current quantum devices are incapable of robust error-correction on long sequences of gate operations, VQAs serve as a near-term bridge for bench-marking real world applications on NISQ devices by taking advantage of classical computation to mitigate noise and coherence times on current devices. Understanding the structure and adaptive nature of VQAs to handle the constraints of near-term quantum computers is integral to achieving fault-tolerance in the long-run~~\cite{shor_fault-tolerant_1997}

The optimization framework for parameterized quantum circuits(PQCs) used in VQAs hold many parallels to variational techniques from classical machine learning for neural networks where gradient methods are used in the context of optimization problems to minimize non-linear functions~~\cite{bottou_large-scale_2010}. In supervised learning models for classical neural networks, a goal of optimization is to learn the weights of node mappings $\mathbb{R}^n\xrightarrow{} \mathbb{R}^k$ between layers using derivative based cost functions to calculate the loss between parameter updates. It is an open area of research to identify suitable quantum problems where complex correlation between data is classically hard and leads to better model  expressibility over classical neural networks as deep learning models can have on the order of 100 billions of parameters leading to costly training schemes~~\cite{lecun2002efficient}. Though PQCs and quantum neural networks are used interchangeably in literature,  Quantum neural networks are an application of PQCs which leverage a quantum-state encoding function and classical optimization loop~~\cite{abbas_power_2021}.  \newline 
However, both PQCs and neural networks suffer from vanishing gradients due to the increasing complexity of the parameter space~~\cite{de_palma_limitations_2023}. This phenomenon is known as a barren plateau in quantum literature and is one of the most prevalent challenges for achieving scalability for variational quantum algorithms. As our system size increases, the loss landscape becomes increasingly non-convex and makes it harder to find an optimal parameter set for our objective function, rendering gradient-based optimization unusable in the context of the exponential dimensionality of the Hilbert space~~\cite{wiersema_here_2023}. Various factors lead to the emergence of barren plateaus in PQCs including include global cost functions, Haar random parameter initialization, unstructured ansatz, information scrambling, noise, and highly entangled circuits~~\cite{harrow_random_2009, hunter-jones_unitary_2019, wang_noise-induced_2021, patti_entanglement_2021, uvarov_barren_2021, napp_quantifying_2022}. If the issue of barren plateaus can be circumvented in PQCs, then classically intractable components of variational problems may be successfully implemented on NISQ hardware in the limit of lower coherence time and noise to serve as initial prototypes for hybrid quantum-classic applications~~\cite{zhu_training_2019}. \newline

In this work, we explore various methods for both mitigating and detecting the onset of barren plateaus, the relation between barren plateaus classically vanishing gradients in classical machine learning, and build the background through mathematical formalisms of the genesis for barren plateaus and gradient-based optimization for PQCs. Furthermore, we explore the influence of circuit geometries on barren plateaus. 
 
\section{Background and Definitions}\label{section:background}
\subsection{Parameterized Quantum Circuit} \label{sec: pqc}
A parameterized quantum circuit is a quantum algorithm that depends on fine-tuning free parameters and quantifies how expressive the chosen family of circuits is~~\cite{wecker_towards_2015}. We can express a PQC,  $U(\theta)$, as: 
\begin{equation} \label{eq: pqc equation}
 U\left( \theta \right) =\prod ^{1}_{i=L}W_{1}U\left( \theta _{1}\right)W_{2} U\left( \theta _{2}\right) \ldots \left( \theta _{L}\right)W_{L}
\end{equation}

Where $U(\theta) =e^{-i \theta V}$ to represent real-valued parameters, $V$ is a hermitian operator, and $W$ be fixed unitaries such as entangled gates or CZ gates. The optimization problem for the PQC takes the form:
\begin{equation}\label{eq:unitary}
    \theta^*=\underset{\theta}{\operatorname{argmin}} C(\theta) 
\end{equation}

Generally, the cost function is modeled as a function of the expectation value of an Observable $\hat{O}$ value to quantify the evolution of the trial state averaged over observed outcomes. This takes on the form:
\begin{equation} \label{eq:cost}
    C(\theta)= f( \bra{0} U^{\dagger}(\vec{\theta}) \hat{O} U(\vec{\theta})\ket{0})
\end{equation}

and the choice of $f$ depends on the specific problem of interest for the PQC. To find the gradient of the objective function, we take the partial derivative of each parameter with respect to the expectation value of our trial state. Thus, the gradients of cost function can be described as:
\begin{equation}\label{eq:partial}
    \partial _kC \equiv \frac{{\partial C({\boldsymbol{\theta }})}}{{\partial \theta _k}}
\end{equation}

the gradients of each unitary and time-dependent parameter sets are fed to a classical optimizer which updates the parameters based on a gradient-descent strategy~\cite{cerezo_variational_2021}.
\subsection{Circuit Ansatz}
A circuit ansatz describes the gate sequence that VQAs use in the optimization subroutine~~\cite{du_quantum_2022}. It can be thought of as the circuit geometry that prepares the trial state through iteratively updating the parameters of the gates. We want the unitary ansatz as described by \ref{eq:unitary} to evolve to a target unitary $V$ such that $V^{\dagger} U = I$, meaning we have found the appropriate set of parameters for our objective function from our linear transformation of the input state. These ansatz can be problem-specific, hardware-efficient, problem-agnostic, or dynamic~\cite{nakaji_expressibility_2021, napp_quantifying_2022}. Circuit geometries as described in \ref{sec:geometry} play in important role in cost landscapes encountering barren plateaus.
\subsection{Unitary groups} \label{sec:unitary-group}
A unitary matrix is an $N \times N$ such that $U^{\dagger}U = I$ or in other words, is orthonormal in $\mathbb{C}^{n}$ \label{def:ugroup} with $N^2$ free parameters~\cite{itzykson_unitary_1966}. Thus, our goal with finding a parameter set with a circuit ansatz it to satisfy the orthogonality condition for the parameterized gates as quantum operations must be reversible. From the definition of orthogonality, the eigenvalues of unitary matrices,$\lambda$,  thus lie on the unit circle, so $\lambda = e^{i\theta}$ for $\theta \in \mathbb{R}$. 
$U(N)$ refers to a group of $N \times N$ matrices in $C^n$ such that the inner product (norm) is preserved. Therefore, our variational circuit finds a solution space of unitary parameters from the entire possible distribution space of $U(N)$. Since the unitary group is a manifold, we can find curves through points in $U(N)$, allowing us to approximate gradient functions using a tangent space~\cite{isaev2002effective}. \newline

A special unitary group, $SU(N) \subset U(N)$, is a special subgroup such that the determinant of every element $a \subset SU(N)$ = 1 ~\cite{gamburd1999spectra}. The number of free parameters for an element in the $SU(N)$ group is $N^2 - 1$. Since global phase has no influence on measurement probabilities, all quantum gate operations can thus be represented by a matrix $A \subset SU(2^N)$.  \newline

\textbf{Derivation.} As per definition of the unitary group (\ref{def:ugroup}), a complex matrix ha $n^2$ free parameters, and each complex value has 2 real parameters leading to a total of $2n^2$ parameters. Since diagonal elements in hermitian matrices must be real and the determinant must be one, We pick vectors for each row such that each vector is $\perp$ to the span of the previous rows.\newline

The $SU(N)$ group is of particular interest for circuit ansatz due to the below theorem:\newline
\begin{theorem}
[Solovay-Kitaev]. Given a set of elements in $SU(N)$ that generates a dense subset, then it is possible to find approximations for any element of
$SU(N)$ with short sequences of elements of the given set.
\end{theorem}
In essence, this theorem guarantees that by using a universal gate set for our PQC, the error is arbitrarily reduced to approximate a unitary as we increase the number of gates~\cite{dawson_solovay-kitaev_2005}. Though this theorem guarantees that we can find a short sequence of gates that scales $\mathcal{O}(\log(1 / \epsilon)$ with the error to estimate a single unitary gate, it is an open question as to how circuit geometries can take advantage of the underlying symmetries to produce an efficient decomposition of gates for a PQC. Furthermore, it is worth exploring a mapping for $SU(N)$ for Haar measures to prompt exploring symmetric circuit ansatz for PQCs to take advantage of problem-specific structures in data~\cite{meyer_exploiting_2023}.

\subsection{Unitary Ensemble} \label{sec: unitary-ensemble}

To find a parameterization for a PQC encoded problem by some $U(\theta)$ from\ref{sec: pqc} requires parameters to be sampled from an underlying probability distribution within $U(N)$. Though matrix entries themselves are random variables, we can generalize a measure of uniformity by equivalently considering a probability measure on sets of matrices from the unitary space ~\cite{riser_power_2023}. PQCs encode a unitary evolution, so a unitary ensemble is a collection of matrices equipped with a probability measure that is left and right(translational) invariant under symmetries and conservation laws of quantum systems described as such~\cite{mele_introduction_2023, hunter-jones_operator_2018}: 
\begin{equation}
    \forall V \in U(N),~UV \simeq UV \sim \mu_{\rm Haar} 
\end{equation}

Symmetries in quantum systems describe the physical properties and constraints of the problem encoded by a circuit ansatz for a PQC(i.e spin systems, lattice Hamiltonian)~\cite{meyer_exploiting_2023}. Thus, quantum symmetries are a generalization of group algebras which preserve certain properties of the system and give rise to multiple classes of unitary ensembles. \newline

\textbf{Circular Unitary Ensemble} Of particular interest is the circular unitary group. This ensemble represents a compact lie group that is invariant under all $n \times n$ matrix transformations~\cite{dakovic_circular_2016}. This measure is described as the haar measure discussed in \ref{sec:haar}. The eigenvalues are distributed evenly along the unit circle in the complex plane; this ensures that the states produced by PQCs can sufficiently explore the state space~\cite{gross_evenly_2007}. The joint probability distribution of eigenvalues for an $N xN$ matrix given by Dyson [cite] initially for the CUE is:
\begin{equation}
    P(\psi_{1}, \ldots, \psi_{N}) = C_{\beta} \prod_{i <j} | e^{i\psi_{i}} - e^{i\psi_{j}}|^{\beta}
\end{equation}
where $C_{\beta}$ is a normalization constant, $e$ contributes to the decay or larger eigenvalues, and the absolute value term quantifies the repulsion between pairs of eigenvalues~\cite{pozniak1998composed}. This statistical-mechanical matrix ensemble allow us to explore different circuit characterizations for physical systems under the constraint of time-reversal and implies that the statistical behavior of energy levels becomes more predictable when averaged over various complex systems.

\subsection{Random Unitaries} \label{sec:ran_un}

A random unitary for an $n$ qubit gate is a unitary matrix sampled from probability distribution $\mathcal{D} \subset U(N)$. In quantum systems, we are interested in understanding how well an ensemble approximates the unitary group and thus, its inherent randomness~\cite{zyczkowski_random_1994}. The frame potential in  gives us an estimate of the 2-norm distance between an ensemble $\varepsilon$ and the Haar measure of $U(N)$. Random unitaries and circuit symmetries give rise to the phenomenon of information scrambling of quantum systems during evolution and entanglement properties between parts of a system~\cite{bengtsson_frame_2008}.

In a random unitary for PQCs, all parameters are considered to be i.i.d. from underlying distribution and thus, a random matrix describes the joint probability distribution $\varepsilon$.

In previous sections, we noted the circular unitary ensemble was equipped with the haar measure; notably, for unitaries to be truly haar random involves a large amount of degrees of freedom making the simulation unphysical; more physical is the random circuit model where random local quantum gates constrain the degrees of freedom due to local interactions that model graph connectivity, leading to more efficient use of local gates and qubits~\cite{brandao2016local}. Approximations have led to t-designs \ref{sec:t-design} that resemble the haar distribution up to t moments with lower computational complexity, but approximating t-designs is an open challenge as barren plateaus are prevalent due to regions of parameter spaces with gradients near zero~\cite{harrow_approximate_2023}.

\subsection{Haar Measure} \label{sec:haar}

\begin{equation}
    {\int}_{\mathcal{U(N)}} {d\mu (U)f(U)} = {\int} d\mu (U)f(VU) = {\int} {d\mu (U)f(UV)}
\end{equation}

 A measure describes how items are distributed in a mathematical space such as a probability distribution for $U(N)$ ~\cite{mele_introduction_2023}. We care about completeness of measurement for state evolution; specifically, completeness in PQCs means that the unique number of measurements must fully represent the entire space of possibilities for an $n$ qubit system. A Haar measure for $U(N)$ describes a left/right invariant measure for sampling points from a uniform distribution. For example, a Hilbert space of dimension 2 means we sample points from the surface of the bloch sphere. Expressiveness of a PQC can be quantified by the potential of the circuit to fully explore $U(N)$ and its distance to the uniform Haar distribution~\cite{du_expressive_2020}. The more expressive a circuit, the closer its distribution resembles a haar measure; however, largely expressive circuits lend themselves to being prone to barren plateaus~\cite{haug_capacity_2021, holmes_connecting_2022}. Circuit expressivity is discussed in \ref{sec: expressibility}.
 
\subsection{Unitary t-designs} \label{sec:t-design}

t-designs can be understood as the probability distribution over the unitary ensemble, $\{p_{i}, V_{i}\}$ for the space of unitaries that a circuit ansatz can reach  such that it is hard to sample from the probability distribution with a classical device~\cite{haferkamp2022random}. A t-design matches the Haar distributing up to the t-th moment, meaning that the PQC has produced a sufficiently random circuit. Thus, the relationship between the Haar measure and t-design can be described as:

\begin{equation}
    \mathop {\sum}\limits_i p_iV_i^{ \otimes t}\rho (V_i^\dagger )^{ \otimes t} = {\int} d\mu (U)U^{ \otimes t}\rho (U^\dagger )^{ \otimes t}.
\end{equation}

where the sum of polynomial functions of at most degree t for $U$ is the same as integrating over the domain of $U(N)$. Another way to intuitively understand this is that t-design means we need t-copies of our trial state to be simultaneously sampled and check its distinguishablitiy from the Haar distribution. Thus, the trial state generated by the PQC may produce approximate t-designs that are pseudo-random enough to capture the  expressibility of the full parameter space for $U(N)$~\cite{gross_evenly_2007}. In essence, t-designs mimic the Haar measure but require less resources to implement. \newline

\textbf{2-Design} From the definition of t-designs from \ref{sec:t-design}, it is sufficient to construct a 2-design for an operationally equivalent action of sampling from the Haar measure for the space $U(N)$. Given the wide class of protocols requiring 2-designs, we will explore approximate 2-designs which relax the uniformity constraint to measure a finite subset of $U(N)$ ~\cite{harrow_random_2009}. Consider  a set of unitaries with their respective probabilities,
$\mathcal{D} = \{(p_{i}, U_{i}) \}_{i=1 \dots n}$. The maps
\begin{equation}
    \mathcal{G}_{w} = \sum_{i}p_{i}U_{i}^{\bigotimes 2} p ( U_{i}^{\dagger})^{\bigotimes 2}
\end{equation}
\begin{equation}
    \mathcal{G}_{H} = \int_{U}U^{\bigotimes 2} p ( U_{i}^{\dagger})^{\bigotimes 2}dU
\end{equation}
for an operator $p$ within the set of linear operators on the Hilbert space is an  approximate 2-design if:
\begin{equation}
    \| \mathcal{G}_{w} - \mathcal{G}_{H} \| \leq \epsilon
    \label{eq:2-design}
\end{equation}
meaning that the quantum channel approximates the ideal 2-design within a bound of $\epsilon$~\cite{harrow_approximate_2023}. 

\subsection{State Fidelity}
\begin{equation} \label{equation: fidelity}
    \langle F \rangle \equiv \int_{\mathbf{U}(D)} dU Tr [ U | 0 \rangle \langle 0 | U^\dagger \Lambda | 0 \rangle \langle 0 | U^\dagger]
\end{equation}
State fidelity, also known as state overlap, measures how close two quantum states are to each other. Since PQCs often rely on preparing quantum states that represent solutions to certain problems, the success of these algorithms depends on the ability of the parameterized circuits to generate states that are close to the ideal solution states~\cite{zhang_lower_2016}. The Haar average fidelity in Eq. \ref{equation: fidelity} contains a quantum channel, $\Lambda $, describing the unitary operations and the trace distance  quantifies how different $\Lambda $ is from the target operation $U$ when averaged over all possible unitary matrices~\cite{dankert2009exact}. The Haar-averaged fidelity can be related to two standard benchmarks for PQCs: the entanglement fidelity of the quantum channel as well as the gate-fidelity of implementing a target unitary on a noisy quantum device for which many experimental schemes are proposed to estimate the former. \newline

Fidelity can also be expressed as in the context of a ground-state problem when measuring the overlap of an optimized state and the target state:
\begin{equation}
    \mathcal{F} = | \bra{\psi(\theta}\ket{\psi_{ground}}|
\end{equation}

State fidelity is an inherently hard problem lying within the QSZK complexity class; as the density matrix scales exponentially, determining the magnitude of the trace distance is computationally expensive problem even for a quantum computer~\cite{chen2021variational}. Within PQCs, process tomography and variational trace distance algorithms are proposed as verification tools for 
quantum information processing tasks and quantify how well quantum information has been preserved~\cite{levy2021classical}.

\subsection{Frame Potential}\label{sec: frame-potential}

\begin{equation}
    \mathcal{F} = \int_{\mathbf{U}}\int_{\mathbf{U}}dUdV\bra{0}(UV^{\dagger})\ket{0}^4
\end{equation}

Another measure of distance between quantum objects linked to 2-designs in \ref{sec:t-design} is the frame potential. In this case, instead of focusing on the distinguishable of the 2 states, we focus on a measure that quantifies the difference between an ensemble of unitaries with respect to and the Haar measure~\cite{cotler2017chaos}. The frame potential used the trace distance which has an operational interpretation as the maximum success probability of distinguishing two states~\cite{bengtsson_frame_2008}. Different works utilize different p-norms, but it suffices to use the 2-norm as it is a probability preserving operation since we are dealing with complex amplitudes(distributions on a unit circle). $L^{2}$ is also the space of all functions where the integral is finite such that $|| \psi || = \sqrt{\int_{X}|\psi(x)|^{2}dx} $ and since we are dealing with the probability of finding a state within a certain set X, the integral must converge to 1 as it reflects all the possibilities. Looking back at \ref{eq:2-design}, we can then express the $\epsilon$ approximation in terms of the frame potential:
\begin{equation}\label{equation:frame-potential}
    \| \mathcal{G}_{w} - \mathcal{G}_{H} \| = \sqrt{\mathcal{F_{\mathbf{U}}} - \mathcal{F_{\mathbf{Harr}}}}
\end{equation}

\subsection{Barren Plateau} \label{sec:bp}
 Let us formalize a definition for the barren plateau. A barren plateau landscape is one where the average
magnitudes of the cost gradients are exponentially suppressed~\cite{mcclean_barren_2018}. Considering the cost function defined in \ref{eq:cost}, a barren plateau will occur if for all $\theta_l \in  \mathbf{\theta}$, the expectation value of the cost function derivative evaluates to 0:
 \begin{equation}
     \begin{array}{*{20}{l}} {\langle \partial _kE\rangle } \hfill =  \hfill & 0 \hfill \end{array},
 \end{equation}
 
and the variance, or second moment, vanishes exponentially with the number of qubits as:
\begin{equation}
   Var_{\theta}[ \partial_{\mu}C(\theta)] \leq F(n),  F(n) \in \mathcal{O}(1/b^{n}) 
\end{equation}

In connection to Chebyshev's inequality, the probability that the cost function partial derivative deviates from the mean value of 0 (using a standard deviation of k) is upper bounded by:
\begin{equation} \label{eq: derivative-bp}
    \frac{Var_{\theta}[ \partial_{\mu}C(\theta)]}{k^2}
\end{equation}

Thus, if variance for the partial derivative cost function vanishes exponentially, it follows that the probability for a non-zero partial derivative vanishes accordingly. In section \ref{sec:geometry}, we will further define a deterministic concentration that is noise-induced, and a probabilistic concentration that is induced by the degree of circuit  expressibility. \newline

\begin{figure}[h]
    \centering
    \includegraphics{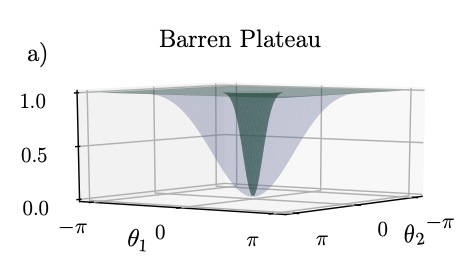}
    \caption{A Narrow Gorge Parameter Landscape. Since the cost values are  exponentially concentrated about the mean value on average, the parameter landscape derivative concentrates about this value leaving the majority of the landscape flat as defined by Eq. \ref{eq: derivative-bp} \.~\cite{arrasmith_equivalence_2022}}
    \label{fig:enter-label}
\end{figure}

We can interpret a barren plateau as a concentration of the cost measure.~\cite{arrasmith_equivalence_2022} This means that the parameter landscape will form a narrow gorge which demonstrates exponential cost concentration about the mean as the system size grows.



\section{Circuit Characterizations and Metrics} \label{sec:geometry}

\begin{figure}[h]
    \centering
    \includegraphics[scale= .8]{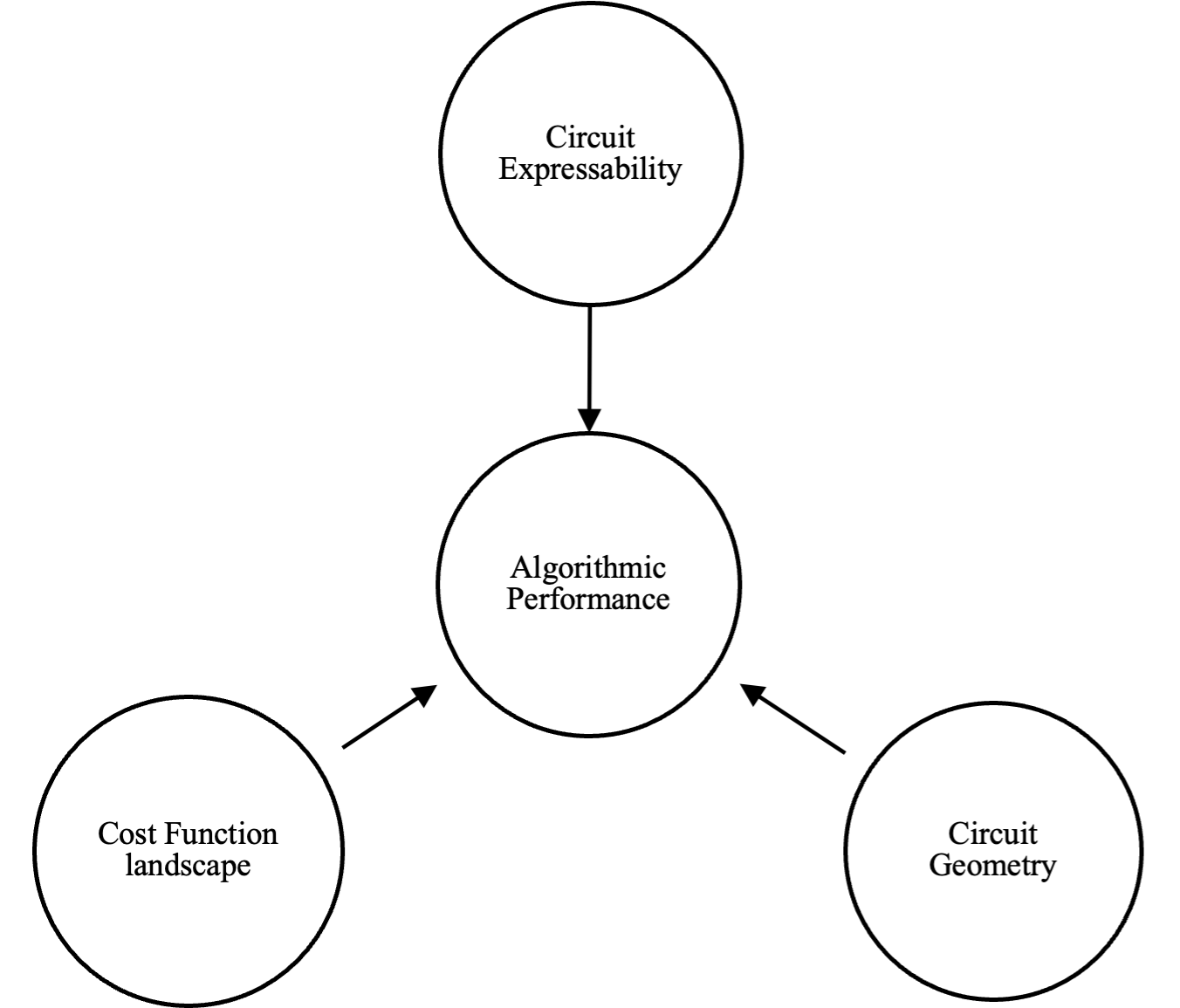}
    \caption{The current landscape of research surrounding PQC design in connection with algorithmic performance for variational tasks. Initial works have studied relationships between cost functions and barren plateaus as well as developed metrics to circuit properties including  expressibility. Most recently, the geometry of a PQC has surfaced as a framework for understanding algorithmic performance.}
    \label{fig:PWC landscape}
\end{figure}

In a parameterized quantum circuit as described by \ref{sec: pqc}, a central question is understanding the best strategy to minimize the objective function \ref{eq:cost}. Naturally, the question arises as what defines "best"? Looking at PQCs from a purely optimization framework obscures the underlying circuit geometry of a variations ansatz~\cite{martin_barren_2023}. Arrangement, connectivity, circuit  expressibility, as well as the relationships between the circuit's parameter encompass the structural aspects of how circuit design influences the behavior of the circuit. Geometry and the cost function landscape plays a crucial role in determining the optimization landscape of the circuit and can impact the efficiency of optimization algorithms used to train the circuit for specific tasks~\cite{milsted_geometric_2018}. Different circuit geometries and their ansatz schemes are crucial to mitigating barren plateaus. In this section, we will take a look at the various ways barren plateaus arise and explore structural paradigms underlying circuit ansatz including chain, alternating, and all-to-all connectivity.
\subsection{Circuit Ansatz} \label{sec:ca} 

\begin{figure}[h]
    \centering
    \includegraphics[scale= .5]{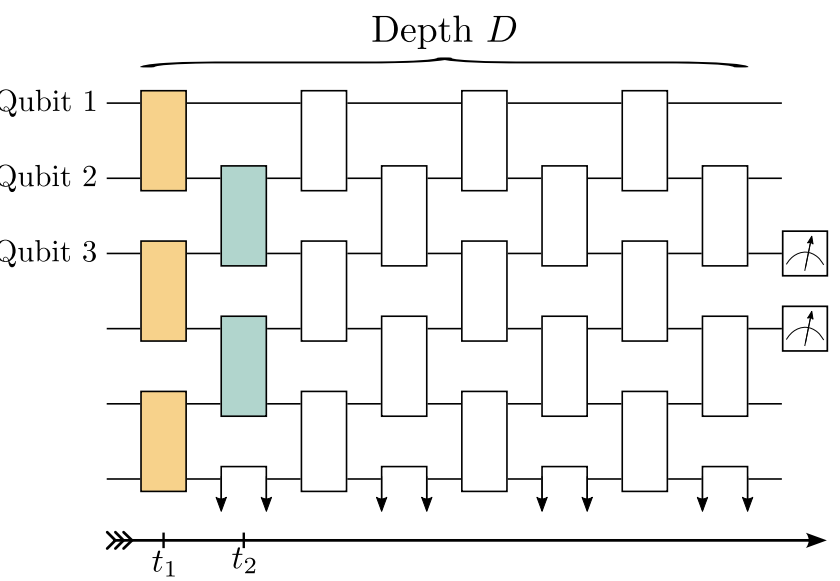}
    \caption{A layered brickwork ansatz scheme. Here, the 2-qubit gates have nearest neighbor connectivity and are used as entangling gates. Parameterized gates are single qubit rotation gates. Dimensionality is the topology of the physical qubits in a chain (1D).~\cite{leone_practical_2022}}
    \label{fig: HEA}
\end{figure}

\subsubsection{\textbf{Hardware Efficient Ansatz}}\label{sec:HEA}
`    A hardware efficient ansatz (HEA) is one in which native gate sets specific to the device at hand are used to mitigate the effect of hardware noise and provide an advantage in gate compilation time. HEAs aim to lower circuit depth and be problem agnostic, leading to a wide array of applications and be an ideal choice for NISQ devices; however, this versatility is a prominent reason why barren plateaus occur for deep HEAs due to their  expressibility~\cite{leone_practical_2022}. Typically, a layout for an HEA is a one dimensional alternating layered ansatz comprised of 2-qubit gates in a brick-like fashion~\cite{nakaji_expressibility_2021}. In some cases, the layer may consist of a layer of single qubit unitary gates followed by entangling gates for all qubits. Looking at the PQC expression in Eq. \ref{eq: pqc equation}, an HEA chooses hermitian operators, $V_{l}$, and unparameterized unitaries, $W_{l}$, from a native gate set determined by hardware connectivity. A single layer, $L$, is defined as a sequence of single rotation gates and 2-qubit gates. Thus, depth, $D$, is the number of repetitions of such a sequence. HEAs have been studied \textbf{(insert ref)} and known to avoid barren plateaus if guarantees can be made about locality and shallowness. HEAs also fail when gates are randomly initialized, which is problematic for large systems~\cite{grant_initialization_2019}. \newline

\subsubsection{\textbf{Physically Motivated Ansatz}}\label{sec:PMA} Another framework for characterizing unitary evolution is as a time-evolution operator rooted in physical systems. These kinds of ansatz start with prior knowledge about the physical system to construct its mathematical form, leading to resource efficiency with a narrower search space. Physically motivated ansatz (PMA) include the unitary coupled-cluster ansatz (UCC) and quantum alternating operator ansatz(OAOA), and Hamiltonian variational ansatz(HVA).\newline

\textbf{2.1: UCC} Looking at Eq. \ref{eq: pqc equation}, the PQC equation for a unitary coupled cluster model typically used for ground state energy problems can be expressed as:

\begin{equation}
    U\left( \theta \right) = \prod_{lm}e^{i\theta_{lm}\sum_{i}\mu_{lm}^{i}\sigma_{n}^{i}} = e^{T-T^{\dagger}}
\end{equation}

Where T is a cluster operator of n-electron excitations with with variational parameters~\cite{anand_quantum_2022}. Thus, $U \equiv e^{T-T^{\dagger}}$ is a time evolution operator with an efficient Hamiltonian. The unitary expression above is parameterized with alternating layers of unitaries. Typically, the circuit structure remains fixed, but variable ansatz do exist where gate may grow or be reduced iteratively as the system evolves. UQCCs are prone to noise-induced barren plateaus affected by the number of layers,and their efficiency often depends on the underlying entanglement spectrum~\cite{wang_noise-induced_2021}. Primarily, they are used to model many-body systems in chemistry.

\subsubsection{\textbf{Quantum Alternating Operator Ansatz}}\label{sec:QAOA}
The Quantum approximate optimization algorithm (QAOA) uses an alternating operator ansatz(QAOA) for finding approximate solutions to combinatorial optimization problems via identifiying an optimal parameter set~\cite{kremenetski2021quantum}. Here, the configuration space includes problem constraints and encodes the set of all possible solutions. The unitary from Eq. \ref{eq: pqc equation} takes the following form:

\begin{equation}
     U\left( \gamma, \beta \right) = e^{-i\beta_{P}H_{M}}e^{-i \gamma_{P}H_{P}}\dots e^{-i\beta_{1}H_{M}}e^{-i \gamma_{1}H_{P}}
\end{equation}

with $ U_{P}(\gamma) = e^{-i\gamma H_{P}}$ as the phase separator family of parameters encoded with the objective function to minimize, and $ U_{M}(\beta) = e^{-i\beta H_{M}}$ as the mixer family that preserves feasible states and provides transitions between feasible states~\cite{hadfield2019quantum}. The QAOA is prone to noise-induced barren plateaus~\cite{wang_noise-induced_2021} \newline 

\subsubsection{\textbf{Hamiltonian variational Ansatz}}\label{sec:HVA}
\begin{figure}[h]
    \centering
    \includegraphics[scale = .4]{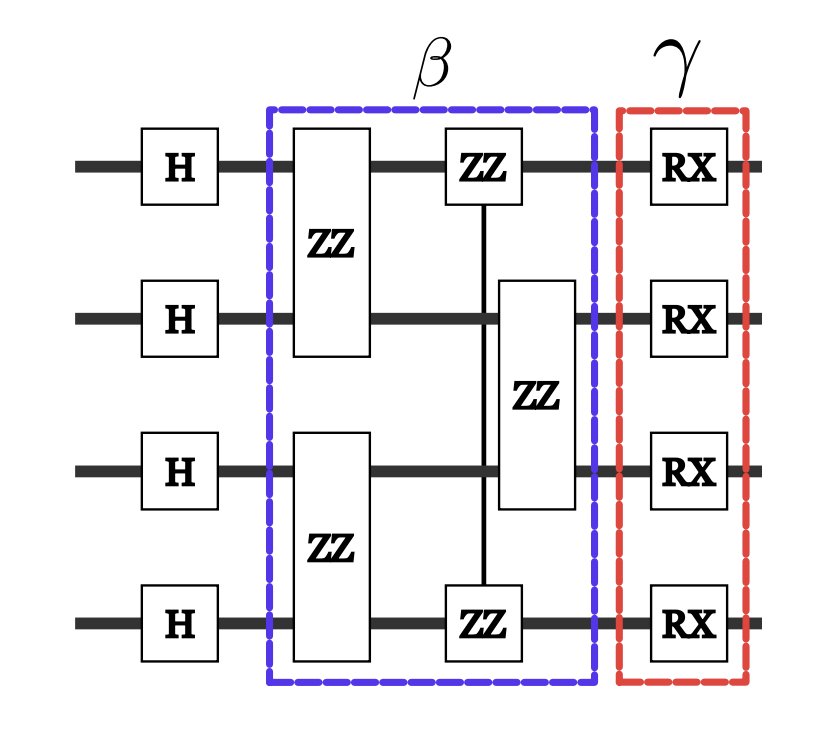}
    \caption{An HVA ansatz. In this example, each layer, d, contains parameterized 2-qubit local ZZ rotation gates followed by CZ entangling gates and a layer of single qubit parameterized X rotation gates~\cite{wiersema_exploring_2020}. In this example, there are 2d parameters for a d-depth ansatz.}
    \label{fig:HVA-ansatz}
\end{figure}

This type of ansatz is a generalization of the QAOA ansatz to more than two non-commuting( i.e, $[H_{s},H_{s'}] != 0$) operators. More genererally, the PQC expression in Eq. \ref{eq: pqc equation} takes the form:
\begin{equation}
     U\left( \theta \right) = \prod_{L}\prod_{j}e^{i\theta_{L,j}H_{j}}
\end{equation}

Where the $H_{jth}$ terms are terms of non-commuting operators. The Hamiltonian is trotterized such that the problem Hamiltonian is encoded as: $H =\sum_{j}H_{j}$. Ideally, the manifold, or set of all possible parameterizations of the wave function, contains the ground state of the problem Hamiltonian that can be reached with optimization schemes. Of importance to barren plateaus in HVAs is the entangling power that prevents these ansatz from reaching fully-formed 2-designs that suffer from vanishing gradients as well as overparmeterization which prevents local minima from forming~\cite{wiersema_exploring_2020}.  \newline

\subsubsection{\textbf{Tensor Network Ansatz}}\label{sec:TNA}
\begin{figure} [h]
    \centering
    \includegraphics[scale=.8]{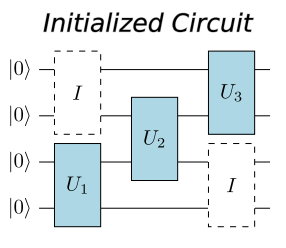}
    \caption{A typical mapping for a 1D tensor network to an MPS ansatz. The staircase pattern is a typical mapping of the local entangling properties of the system. Here, the MPS is used as an approximation solution to pre-train a HEA circuit~\cite{dborin_matrix_2022}. }
    \label{fig:MPS-ansatz}
\end{figure}

Another recently studied ansatz is the tensor network (TNA) based PQC with connections to tensor network structure in classical machine learning. As a general premise,  Tensors characterize the dimensions of vectors and matrices via indices~\cite{noauthor_tensor_nodate}. A tensor network contraction involves pairing and summing indices (regarded as the dimension) according to the network connectivity. When two tensors share an index they are contracted by multiplying the corresponding elements and summing over the index. For a one dimensional lattice system(1/2-spin chain), the quantum state as a matrix product state (MPS) is represented by:

\begin{equation} \label{equation:mps_equation}
    \ket{\psi}  = \sum_{s_{1} \dots  s_{L}} A_{s_{1}}^{1} A_{s_{2}}^{2} \dots A_{s_{L}}^{L} \ket{s_{1} \dots s_{L}}
\end{equation}
Where each $A_{s_{i}}^{i}$ is a rank-3 tensor with dimension $(m_{i} \times m_{i+1} \times d_{i})$ with equality between neighboring states for dimensionality and $d_{i}$ which quantifies the dimension of each qubit~\cite{liu_presence_2022}. Depending on closed or open boundary conditions (indicated as a line for each index), the first and last tensors, $A_{s_{i}}^{i}, A_{s_{L}}^{L}$ may or may not have the same $m_{i}$ to form a ring. Lets explore a diagram of the MPS for a system with 3 site closed boundary condition MPS with a bond between each neighbor:

\begin{figure}[h]
    \centering
    \includegraphics[scale =.3]{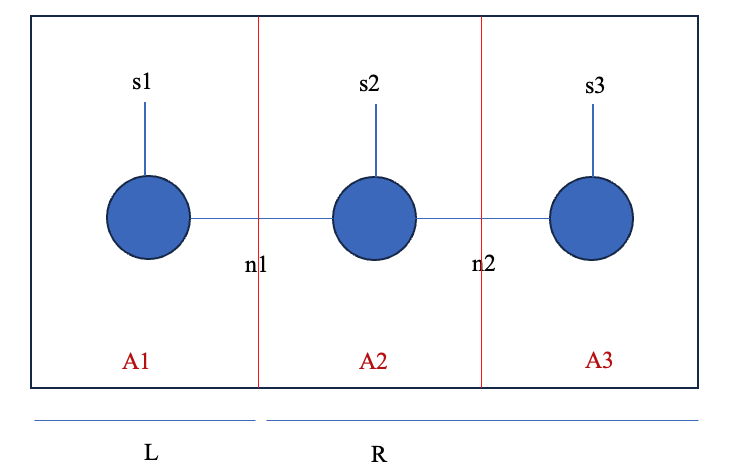}
    \caption{A 3-site closed boundary system with contraction between neighboring sites. Each $A_{s_{i}}^{i}$ is associated with its respective site. A cut of the system may be taken at the L and R subsystems to characterize the Schmidt decomposition for a bipartite system.}
    \label{fig:MPS-diagram}
\end{figure}

The A matrix represents the contraction on each site for the same dimension, $n_{i}$. Connecting back to eq.\ref{equation:mps_equation}, this would scale as $\mathcal{O}(d^{L})$ since each A matrix has $d^{2}$ elements. However, the contraction indices may have redundancy (i.e, dependent vectors). To reduce the complexity, Singular value decomposition(SVD) has been proposed as a way to reduce the complexity for the state computation. Physically, ground states have an eigenvalue spectrum that is not highly entangled with the rest of the system. Thus, our SVD may take the A matrix to a  $(m_{i} \times M \times d_{i})$ operation where M is the bond dimension characterized by the spectral decomposition of the eigenvalues when we consider a bipartite system (i.e, taking a cut of the system as in Fiq. \ref{fig:MPS-diagram}~\cite{gamburd1999spectra}. This has the effect of reduced the rank of the matrices, leading to a parameter scaling with $\mathcal{O}(LdM^{2})$ complexity that is linear in the number of qubits and polynomial in the bond dimension~\cite{martin_combining_2023}. In sec \ref{sec: entanglement}, the consequence of volume and area law entanglement in \ref{sec: entanglement} will be introduced in connection to the consequences on barren plateaus.\newline

Many recent PQCs mappings have been proposed based on tensor network models such as MPS, qMERA, and the Tree Tensor Network structure~\cite{martin_barren_2023, zhao_analyzing_2021}.  qMERA, for instance, has been used as a circuit architecture to study long-range correlations with feasible numbers of qubits.~\cite{anand2023holographic}. Another application of MPS ansatz is holographic quantum simulation using compressed approximations of ground states with lower dimensional cross-sections for qubits to demonstrate interactions between higher dimensional systems ~\cite{PRXQuantum.3.030317, PRXQuantum.3.030317}. Both of these applications of MPS states attempt to take advantage of the polynomial reduction in qubits using the entanglement entropy and use fewer parameters in the pqc. As the bond dimension (degree of entanglement) increases, mapping tensor networks to PQCs and requires more depth and connectivity( swap gates). Typically, most TNA will form a product state of the form $\ket{\psi_{1} \otimes \dots \ket{\psi_{n/m}}}$ with m qubit blocks in the ansatz.\newline

While a low bond dimension is advantageous for capturing short-range entanglement, it might not be suitable for systems with long-range correlations or more complex structures that are prone to the barren plateau phenomenon due to high-depth circuits with vanishing gradients~\cite{haegeman_geometry_2014}; however, this is an active area of research as noted in the works above. 

\subsection{Circuit Metrics}
In the previous section, we laid the foundation for the different types of PQC geometries. With an understanding of different circuit ansatz schemes and their parameterizations to gate connectivity, a central question arises regarding how to quantify the performance of various classes of PQCs on current NISQ devices. Circuit metrics have been developed to draw a connection between the underlying circuit properties and conditions that induce barren plateaus. Thus, we start with another set of definitions for circuit  expressibility, trainability, and entanglement in connection with barren plateaus. \newline

\subsubsection{\textbf{ expressibility}} \label{sec: expressibility}
\begin{figure}[h]
    \centering
    \includegraphics{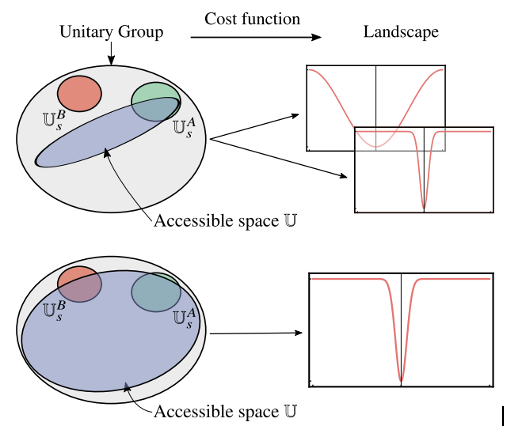}
    \caption{Here $U_{B}$ and $U_{A}$ are two solution spaces to distinct problems. The blue circle in the top figure is a low-expressive ansatz, where as the bottom figure is a highly expressive ansatz for two different PQCs.~\cite{holmes_connecting_2022} }
    \label{fig: expressibility}
\end{figure}

In \ref{sec: unitary-ensemble}, We stated that our PQC is encoded by an underlying unitary ensemble that induces a measure on the unitary group(Def. \ref{def:ugroup}) to restrict our search space to a subspace of $U(N)$. Thus, every ansatz is capable of producing a unique distribution quantified by its ability to generate states that are well representative of the Hilbert space~\cite{du_expressive_2020}. The degree to which that subspace is capable of exploring the full unitary group quantified by the haar measure in Def. \ref{sec:haar}. Eq. \ref{eq:2-design} is thus a measure of  expressibility of the degree that a particular PQC matches a true 2-design~\cite{holmes_connecting_2022}. Figure \ref{fig: expressibility} provides a visual interpretation. 

Notice also how uniformity plays a role in the  expressibility of an ansatz. Non-uniform expressitivity is associated with problem-inspired ansatz such as those in \ref{sec:PMA}. Based on the physical constraints of the system at hand, the subspace the ansatz explores may become non-uniform as shown in the first image of \ref{fig: expressibility} and demonstrate completeness for certain classes of problems while excluding others. On the other hand, an HEA \ref{sec:HEA} is complete and expressive. Therefore when Eq.\ref{eq:2-design} is small, the ansatz is more expressive and demonstrates statistical properties close to a haar-distirbuted state. However,  expressibility does not guarantee trainability. Consider the cost function landscapes from Fig. \ref{fig: expressibility}. For highly expressive ansatz, the cost function landscape of Eq. \ref{eq:cost} is inherently flat; for non-expressive ansatz, both small and large gradients are present for Eq. \ref{eq:cost}. \newline

To formalize a definition of  expressibility, let's take examine Eq. \ref{eq:partial}. When we take the partial derivative of a particular unitary in a PQC described as Eq. \ref{eq:unitary} \, we leave a left and right sequence of unitaries  such that $U(\theta) = U_{L}(\theta)U_{R}(\theta)$, Where, 
\begin{equation}\label{eq:left-and-right}
    U(\theta)_{L} = \prod_{i=1}^{k}U_{j}(\theta)W_{j}, \texttt{and    }  U(\theta)_{R} = \prod_{j=k+1}^{L}U_{j}(\theta)W_{j}
\end{equation}

describe the partition. If either $U(\theta)_{L}$ or $U(\theta)_{R}$ form a 2-design (Eq. \ref{eq:2-design}) such that the accessible space is $U(N)$, then we arrive at a variance of the partial derivative vanishing exponentially with the number of qubits as defined by \ref{sec:bp}  describe generalizations of the variance bound when these partitions form $\epsilon$-approximate designs~\cite{holmes_connecting_2022}.
\subsubsection{\textbf{Trainability}}

If lower expressively does not guarantee our solution state to be within the parameter space or contain cost function landscapes with gradients, how then can we quantify when a general circuit will be trainable? From this notion, there is a tension between an ansatz being trainable and its optimization ability in terms of the cost function~\cite{holmes_connecting_2022}. But generally, from the formulations from Barren Plateaus  \ref{sec:bp} on the variance and cost function derivative, we can formalize the following conclusions on trainability:
\begin{enumerate}
    \item A perfectly expressive ansatz (unitaries are haar-random) will always exhibit a barren plateau and is untrainable.
    \item an inexpressive ansatz \textbf{may} be untrainable. 
\end{enumerate}

A further discussion on experiments involving varying circuit depth and the scope of the measurement operator can be found in \ref{sec: bp-geom}. These formalizations demonstrate that problem-inspired ansatz \ref{sec:PMA} will be of interest in the future to uncover problems with interesting symmetries in the underlying space lead to the target unitary contained in the inexpressive ansatz region~\cite{anschuetz_beyond_2022}.

\subsubsection{\textbf{Circuit Geometry}} \label{sec:paramater-landscape}
Turning to the last component of algorithmic performance for PQCs and mitigating barren plateaus are geometric structures. Recent research has been done to characterize quantum states in terms of principle fiber bundles~\cite{haegeman_geometry_2014}. This characterization is based on the kahler space equipped with a complex structure, a Riemannian structure, and a symplectic structure. Fiber bundles provides a way to understand how local symmetries in a base space, the manifold,  relate to global symmetries in a total space~\cite{heydari_geometric_2016}. This relates to an area in quantum mechanics known as gauge theories where mathematical operations that depends on an arbitrary smooth functions in spacetime leave the physical degrees of freedom for the global system unchanged. Recent research has been done to formalize definitions of geometric meaning for~\cite{haug_capacity_2021} PQCs.For ansatz schemes, the underlying topology and curvature of the parameterization landscape in a geometric context through principle fiber bundles has implications for how different parameterizations may influence the trainability of the ansatz. \newline 

Circuit  expressibility and entanglement can be reformulated as the relationships between scalar manifolds and concurrence. From Ref. ~\cite{wootters2001entanglement}, Concurrence for a state, $\psi = \sum_{i} c_{i} \ket{a_{i}}$, $c_{i} \in {\alpha, \beta, \gamma, \delta}$ can be defined as:
\begin{equation}
    C(\psi) = 2|\alpha \delta - \beta \gamma|
\end{equation}

Looking for a  pure-state ensemble with minimum average entanglement for a given mixed state is akin to looking  for a set of states that all have the same entanglement or in other words, the same concurrence. \newline

Characterization of the underlying geometry depend on the underlying circuit structure to formally re-parameterize the quantum state for both the degrees of freedom one observer has access to for a given entangled state and the amount of entanglement shared between qubits~\cite{katabarwa_connecting_2022}. Of  important here for PQCs, is how to update the parameters using a better technique than Eq.\ref{eq:cost} in terms of local geometric expressibility as this does not take into account the parameterization lanscape . In essence, is there a better metric that accounts for the geometric landscape in terms of fiber bundles that determines the best way to update gradients as opposed to the standard gradient-based method? The metrics under consideration are the Quantum Information Fisher Matrix(QFIM) and Fubini-study metric. The QFIM is used in quantum estimation theory to quantify the sensitivity of a quantum state to changes in parameters and characterizes how the state's probability distribution changes with variations in parameters~\cite{liu_quantum_2020}.
Derived from this is the Fubini-Study metric in the context of complex projective spaces to capture the geometry of quantum states. It measures the distinguishability between quantum states in a complex projective Hilbert space with a Riemannian interpretation. \newline

Using these metrics, one can then compute the Quantum natural gradient to update the PQC to the most suitable set of parameters for faster convergence to the target unitary. The partial derivative in Eq. \ref{eq:partial} considers the parameter space to be flat. However, when for some unitary parameters, $\theta_{0} =\theta_{1}$, the flat euclidean space considers this to be an indefinite line. In reality, these angles are a singularity in the parameter space and the volumetric properties of the metric must decrease. Since QNG takes this volumetric change into account for faster parameter convergence, many VQA algorithms prefer this derivative-free gradient updating procedure~\cite{yamamoto_natural_2019}. Other metrics to consider the dimension are the: \newline

\textbf{Parameter Capacity} \label{equation:param-cap}
\begin{equation}
    D_{c}= 2^{N+1} - 1
\end{equation}

which quantifies the total number of independent parameters  the quantum state can express in the Hilbert Space.\newline

\textbf{Effective Quantum Dimension} \label{equation:effective-quant-dim}
\begin{equation}
   G_{c} =  \sum_{i}^{M} \mathcal{I}(\lambda_{i}(\theta))
\end{equation}

which is an indicator function on the rank of the Quantum fisher matrix that determines the total number of non-zero eigenvalues, i.e. the number of independent directions in the state space
that can be accessed by an infinitesimal update of $\theta$. \newline

The study of circuit geometry is the most recent direction in analysis of PQCs trainability; much the underlying geometric structure is yet to be formalized for assessment of barren plateaus. Understanding circuit geometry also lends itself to better initial parameterization for convergence. It is an open question whether geometric interpretations can be generalized to qubits with higher dimensional connectivity beyond two qubit gates.

\section{Factors Underlying Barren Plateau Phenomenon} \label{sec: bp-geom}

With an understanding of different circuit ansatz schemes and the underlying circuit metrics to characterize the effects of expressativity, geometry, and entanglement on trainability, we can look at various way that barren plateaus arise in current PQCs algorithms.

\subsection{Cost Function Dependent Barren Plateaus}

The definition of barren plateaus in \ref{sec:bp} is a probabilistic definition for the likelihood that the cost function gradient will concentrate around 0. There is a relationship between the ansatz depth, underlying qubit connectivity, and scope of the cost function  operation that determines this exponentially decreasing upper bound.

\subsubsection{\textbf{Circuit Depth}}

First, lets explore the ansatz depth for a brickwork ansatz as in \ref{sec:HEA}. A circuit is deep when the number of layers grows $ \in \mathcal{O}(poly(n))$, and shallow when $ \in \mathcal{O}(log(n))$. As the circuit depth(L layers) grows, the localized  m-qubit blocks show on an ansatz such as in fig. \ref{fig: HEA} will from a 2-design (see Def. \ref{eq:2-design}) in either $ \in \mathcal{O}((mL))$ or $ \in \mathcal{O}(m^{2}L))$ depending on 1D or 2D qubit connectivity respectively~\cite{cerezo2021cost}. This is an extension to the relationship of the cost function partial derivative defined in  expressibility in \ref{sec: expressibility}. Thus, since m and L have an inverse relationship, we can conclude that m qubit blocks in the ansatz form 2-designs even at shallow depths in its left and right unitary partitions which allow for more ansatz  expressibility. \newline

\subsubsection{\textbf{Global and Local Cost Functions}}\label{sec:cost-bp}
The cost function plays a big role in determining whether a PQC will exhibit a barren plateau depending on if we evaluate the cost function on the entire circuit or a subsystem of the circuit. Thus, a cost function can be either global or local, meaning the operator in Eq. \ref{eq:cost} is resolved differently. In the global case for an ansatz with a fixed partition of the complete Hilbert space into a tensor product of k subsystems with operator takes on the form:
\begin{equation}
    O_{L} = c_{0} \mathrm{I} + c_{1}\sum_{i=1}^{N}O_{i1} \otimes O_{i2} \dots \otimes O_{i\xi}
\end{equation}
with the cost function in Eq. \ref{eq:cost} defined as:
\begin{equation}
    C = Tr[O_gU(U(\theta) \rho U^{\dagger}(\theta)]
\end{equation}

For the global case, the operator acts on the entire system~\cite{uvarov_barren_2021}. Irregardless of the number of layers in an ansatz such as in fig.\ref{fig: HEA}, the circuit will always exhibit a barren plateau, meaning an exponential number of shots is needed to resolve the best gradient update. \newline

In the local case, an operator acts on up to k qubits in a k-qubit subsystem with the form:
\begin{equation}
    O_{G} = c_{0} \mathrm{I} + c_{1}\sum_{i=1}^{N} O_{i}^{X,Y,Z}  \otimes O_{i}^{X,Y,Z}
\end{equation}
with the cost function in Eq. \ref{eq:cost} defined as:
\begin{equation}
    C = Tr[O_{L}U(\theta) \rho U^{\dagger}(\theta)]
\end{equation}

\begin{figure}[h]
    \centering
    \includegraphics[scale=.7]{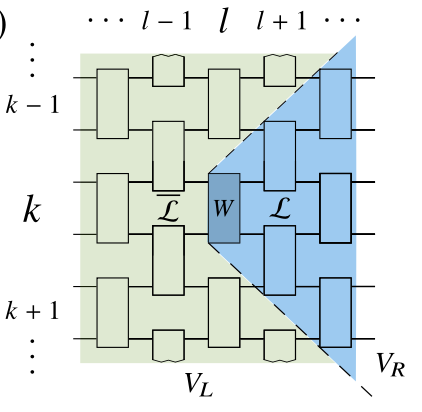}
    \caption{Local Operator light-cone showing the effects of information propagation from the k-th block to the all future entangled gates in the k + cl layers. ~\cite{cerezo2021cost}}
    \label{fig:light-cone}
\end{figure}

For the local operator case, a shallow ansatz such that depth L $ \in \mathcal{O}(log(n))$ will exhibit polynomial vanishing gradients and exhibit barren plateaus with the circuit depth L $ \in \mathcal{O}(poly(n))$ in its cost function gradient. This can be understood by the fact  the entangling gate, $W_{k}$, from the PQC equation in Eq. \ref{eq: pqc equation} is present for each k-qubit system an operator acts on. Each entangling gate propagates the effects of an operator on block k forward in a light-cone pattern since the entangled gates connect blocks at the edges of the m - qubit subsystem~\cite{garcia_barren_2023}. At larger depths, the k-th block operator behaves more like a global operator and induces barren plateaus as the depths increases as seen in Fig.\ref{fig:light-cone}.

\subsubsection{\textbf{Mitigation Strategies}}
Thus for a brickwork ansatz as in \ref{sec:HEA}, the following conditions must hold for a lower bound on the variance of the cost function gradient to prevent barren plateaus~\cite{cerezo_variational_2021}:
\begin{enumerate}
    \item The ansatz must form local approximate 2-designs at each circuit block for its unitary ensemble  to be expressible enough
    \item The cost function utilizes a Hamiltonian with local operators on an m-qubit subsystem
    \item The number of layers must be at least $ \in \mathcal{O}(log(n)$ to allow local 2-designs to form
\end{enumerate}

Of central importance is assessing the scope of the operator to the target problems at hand to weigh a trade-off with trainability and expressiblity. Good parameter initialization strategies play a role for cost-function induced barren plateaus as trainability is better when the local operators are further from the identity~\cite{liu_mitigating_2023}. One such strategy is the identity initialization strategy described below.\newline

\textbf{Identity Initialization Strategy} \label{sec:identity-initialization}
This strategy allows for approximate 2-designs. We saw in Eq. \ref{eq:left-and-right} that when the left and right unitaries in the partial derivative of \ref{eq:partial} for a 2-design, the number of shots needed to update the parameter becomes exponential. Instead we want these blocks to be $\epsilon$-approximate 2-designs as in \ref{eq:2-design}. To mitigate this, randomly selecting some of the initial parameter values while choosing the remaining values in such a way that the result is a fixed unitary matrix, i.e., a deterministic outcome, such that the circuit evaluates to the identity will restrict the effective depth when taking the gradient with respect to any parameter~\cite{grant_initialization_2019}. Some circuits may need an extra entangling layer based on the problem constraints such as the VQE algorithm. A trade-off to consider is if much entangling is needed, the circuit will reach an entanglement-induced barren plateau quicker. Thus, initialization ranges can be specified to the specific problem constraints at hand to be close enough to the target unitary. 

\subsection{\textbf{Noise-Induced Barren Plateaus}} \label{sec:noise-bp}
For NIQS devices and algorithms, it is still an open question as to how the training process of a VQA is affected by noisy hardware. In the presence of noise, qubits decohere, leading to gate errors propagating throughout the ansatz which affects the circuit ever reaching the target unitary. Since every gate introduces noise, there is a relationship the circuit depth and the emergence of a barren plateau as a consequence of noise~\cite{wang_noise-induced_2021}. Previously, we introduced a probabilistic definition of a barren plateau in \ref{sec:bp} in terms of the probability that the cost function ever deviates from the mean value. Hardware noise requires use to turn to a deterministic definition for a barren plateau.  \newline

\textbf{Deterministic Concentration} If the trivial value of the cost function in Eq. \ref{eq:cost} is $\frac{1}{2^n}Tr[O]$ such as the max or min of a cost function, then $C(\theta)$ is $\epsilon$-concentrated error induced by noise channels if for any $\epsilon$ (error) $\geq$ 0 and for any $\theta$:
\begin{equation}\label{eq:deterministic-bp}
    |C(\theta) - C_{triv} | \leq \epsilon
\end{equation}

With a lower bound and upper bound  for $\partial C(\theta)$ as in Eq. \ref{eq:cost} of:

\begin{equation}\label{equation:deterministic-lower}
    L \in \Omega(n), F(n) \in \mathcal{O}(2^{-an})
\end{equation}

Unlike the probabilistic interpretation, the noise-induced barren plateaus(NIBPs) exhibit exponential decay of the gradient itself and induced a flat landscape for the entire parameter space~\cite{leone_practical_2022}. Similarly, $\partial C(\theta)$ decreases exponentially as the number of layers grow linearly with n. the This is unlike in Fig. \ref{fig: expressibility} where the cost function as  expressibility increases becomes more narrow. Thus, there is no global minimum anymore. 

\subsubsection{\textbf{Mitigation Strategies}}
Unlike cost-induced barren plateaus in Ref. \ref{sec:cost-bp}, there is no dependence for noise-free barren plateaus of the parameter choice or locality of the cost function~\cite{wang_noise-induced_2021}. NIBPs are much harder to target, as no layer-wise traing remedies or initialization strategies can resolve the presence of an untrainable landscape due to hardware noise. Whether or noise error-correction strategies can be hep protection information loss is an open question~\cite{martin_combining_2023}.

\subsection{\textbf{Entanglement-Induced Barren Plateaus}} \label{sec: entanglement}

Entanglement is one of the properties that underlies many-body systems who exhibit correlations that may be inherently non-local. As we saw in the MPS in \ref{sec:TNA}, the bipartite cut of the system attempts to quantify the degree to which site L is entangled with the rest of the system, R. Thus, we can use a reduced density matrix to trace-out the two subsystems as such:
\begin{equation}
    \rho_{L}(\ket{\psi} = Tr_{R}(\ket{\psi}\bra{\psi})
\end{equation}
where we can re-write the quantum state in its Schmidt decomposition with $\bra{\psi_{A}^{i}} \ket{\psi_{A}^{m}} = \delta_{im}$ and similar orthogonality condition for B as follows:
\begin{equation} \label{equaiton:schmidt-decomposition}
    \ket{\psi} = \sum_{i}\lambda_{i}\ket{\psi_{A}^{i}} \otimes \ket{\psi_{B}^{i}}
\end{equation}
We can look at the entanglement of a subsystem with the Vonn Nuemann Entropy:
\begin{equation}
    \mathcal{S}(\rho(L)) = -Tr[\rho(L)\log\rho(L)] = \mathcal{S}(\rho(R))
\end{equation}

which is the same irregardless of the subsystem to demonstrate complementary for pure states and holds invariance under unitary transformations. Entanglement is relevant study of quantum phase transitions. When a phase transition at absolute zero temperature  occurs due to varying some system parameter of the Hamiltonian known as the order parameter, the ground state energy that characterized the system may change and exhibit different behaviour due to a change in underlying symmetries~\cite{wiersema_measurement-induced_2023}. Gapped and gapless Hamiltonian are distinguished by their energy gaps between the ground state; gapped Hamiltonians, marked by a finite energy gap between ground and excited states, characterize transitions between ordered and disordered phases, while gapless Hamiltonians, where no energy gap exists, govern critical points with scale-invariant behaviors~\cite{hastings2006solving}. Of particular interest is the spectral decomposition of the ground state. Understanding whether a system is gapped or not by its spectral decomposition leads to different scales of correlation decay which for our purposes quantifies the amount of entanglement in a system. This in turn, governs whether or not a barren plateau arises due to the degree of entanglement within the system~\cite{wiersema_exploring_2020} For VQAs we can characterize systems by either area-law entanglement growth or volume law entanglement growth~\cite{marrero_entanglement_2021}. \newline

\begin{figure}[h]
    \centering
    \includegraphics[scale = .5]{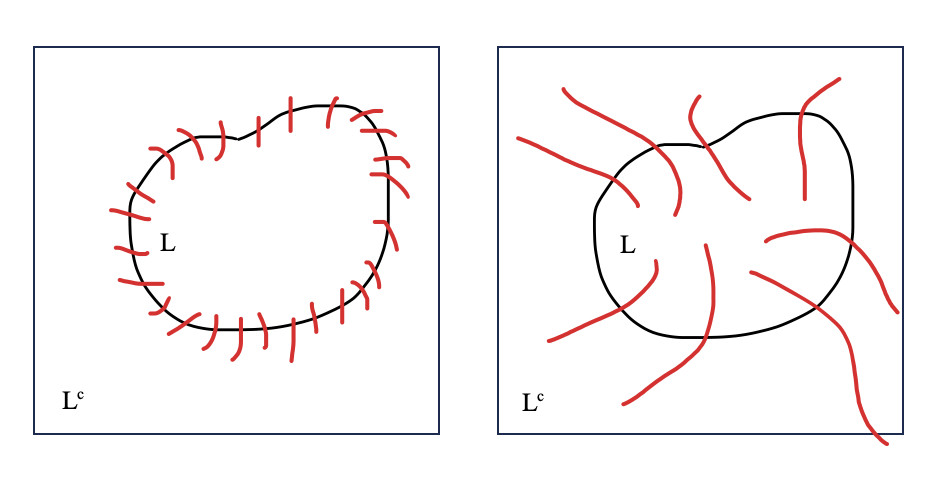}
    \caption{On the left, the growth of entanglement with area law scaling with the boundary of the subsystem. Pictured on the right, the growth of entanglement with volume law scaling with the growth of the subsystem.}
    \label{fig:entanglement_entropy}
\end{figure}

For a state $\ket{\psi}$ in a bipartite Hilbert space described as Eq. \ref{equaiton:schmidt-decomposition} with N qubits and k eigenvalues with a Hilbert subsystems $L^{c}$ and $L$ the area and volume laws are as follows:

\textbf{Area Law Entanglement}
\begin{equation}
    S_{L}\propto |\delta L|
\end{equation}

\textbf{Volume Law Entanglement}
\begin{equation}
    S_{L}\propto |L|
\end{equation}

This relationship quantifies the degree of information scrambling from localized regions to the rest of the system~\cite{li_quantum_2018}. Information scrambling arises from operators acquiring more non-identity basis as well as a growing number of terms required to characterize the operator. 1D chain Hamiltonians with nearest neighbor interactions contain a spectral gap for the ground state that when large enough demonstrates fast convergence to the ground state which implies weak correlations with the rest of the system. ~\cite{hastings2007area} shows that for a d-dimensional chain with spectral gap of $\delta$, the entanglement entropy is bounded by a constant as $e^{\mathcal{O}\frac{log(d)}{\delta}}$. This means we can find an efficient description of the ground state in polynomial time. For 2D systems, more underlying assumptions about the system are needed to ensure area law convergence as no formal proof exists that 2D systems exhibit area law scaling in the general case. 

\subsubsection{\textbf{Random Circuits and Information Scrambling}}
Drawing many sets of random parameters, $\theta_{i}$ is akin to drawing many independent Gaussian distribution. As we increase the sample size to N, the variance of the sample mean decreases proportionally as $\sigma_{\mu} = \frac{\sigma^{2}}{N}$. Thus, Eq. \ref{eq:cost} approximates the mean with low sample variance. This randomness is what drives the concentration of cost function and leads to narrow gorges rather than the large dimensionality of the parameter space. This intuition is why we approximate unitaries as 2-designs in \ref{eq:2-design} when they match the haar measure (Ref. \ref{sec:haar}) since its a measure of uniform randomness~\cite{holmes_connecting_2022}. \newline 

However, circuit connectivity as shown by \textbf{[insert paper]} is what modulates the spread of entanglement through a circuit, as the amount of cost-function connected qubits influences the entropy. Locally entangled gates for random circuits generate entanglement at their boundary at a rate that scales at a volumetric rate and entanglement growth is hard to check on NIQS devices~\cite{hunter-jones_unitary_2019}. We have to be mindful of partitioning the PQCs measurement qubits in such a way that the expressbility is limited by unrecoverable disentangling partitions. Since measurements alter the circuit entanglement and training induces parameter bias, we only assume cannot assume uniform randomness at initialization~\cite{wiersema_measurement-induced_2023}. Otherwise, that implies an inherent parameter structure for sufficient randomness in all PQCs. In fact, our barren plateau \ref{sec:bp} assumptions sufficient randomness throughout, when it is dependent on trainability procedures. Since barren plateaus are believed to preclude the training process, capitalizing on thermalization process throughout a circuit to modulate growth is paramount to larger PQCs models~\cite{srednicki1994chaos}.

\subsubsection{\textbf{Mitigation Strategies}}

\textbf{Weak Barren Plateaus} In connection with PQCs, it is a question of interest then how measurement perturb the entanglement of the system. one of the proposed strategies is a diagnosing a weak barren plateau before the onset of a barren plateau. ~\cite{sack_avoiding_2022} showed formalized the definition of a weak barren plateau using the second reyni entropy defined as:
\begin{equation}\label{eq:second-reyi-entropy}
    S_{2} = -lntr\rho^{2}
\end{equation}
where for a k-qubit subsystem of the Hamiltonian, $\rho_{A}$ and the maximum entanglement for a k-qubit subsystem, $S_{Page}$ A weak barren plateau is such that:
\begin{equation}
    S_{2} = -lntr\rho_{A}^{2} \geq \alpha S_{Page}, \space \alpha \in [0,1]
\end{equation}

A weak barren plateau precludes the onset of a barren plateau definitely with circuit depth at a later point, so this procedure allows the training process to be restarted with a smaller parameter update value.\newline

In all, performing local measurements disentangles the subsystem from the rest of the state and is a method to control the entanglement growth. When a PQC demonstrates volumetric entanglement growth, a barren plateau landscape arises leading to untrainability for a circuit ansatz. therefore, mid-circuit measurements, circuit pre-training, entanglement regularization to add terms to the cost function, and initial partitioning play on the entanglement entropy to better mitigate barren plateaus ~\cite{patti_entanglement_2021, wiersema_exploring_2020, skolik_layerwise_2021} .


\section{Barren Plateaus in Classic Machine Learning} 
Machine learning took an interest in quantum from curse of dimensional where for large inputs, a high number of training sample as needed to extract the desired information. Essentially, training large inputs becomes unfeasible classically especially in the context of many body simulations. PQCs hope to take advantage of intrinsically quantum properties such as entanglement and superposition to make training huge data sets feasible~\cite{sharma_trainability_2022}. Quantum simulations even with a small number of qubits need an exponential amount of memory. However, In the classical world, there are models that are highly expressive and don't suffer from vanishing gradients~\cite{zhang_understanding_2017, raghu2017expressive}. In the quantum sense barren plateaus are much more of a fundamental issue due to the number of measurement shots needed to vary the cost function to account for flattened landscapes, as this is a resource intensive process and lends to expressive or inexpressive ansatz being untrainable~\cite{mcclean_barren_2018}. In this section, we will cover some of the differences between key components of classical neural network(NN) models and parameterized quantum circuits.

\subsection{Classic Network Representation}
Classical neural networks(NN) were the inspiration for the parameterized quantum circuit models seen here. In fact, the breakdown of a PQC in \ref{sec: pqc} draws much inspiration from the structure of a deep neural network~\cite{roberts_principles_2022}. \newline

\textbf{Input Function}
Classically, the input function is a sum of each sample with a weight and bias term. We feed it into the input layer for a multivariate function with multiple inputs represented as a columns in a matrix X as:
\begin{equation}\label{eq:hidden-layer}
    f_{in} = w_{1}*X + b_{1}
\end{equation} 

The input layer for the initial data can be encoded in a quantum circuit via an encoder circuit or as initial gate operations on the circuit that gets encoded as part of the unitary expression as:
\begin{equation}
    \ket{\psi_{in}} = \otimes R(x_{i})\ket{0}^{n}
\end{equation}

\textbf{Hidden Layers} The weights of the nodes to in each hidden layer of a NN are modeled as such and contain a bias term, where for multiple layers represented a multi-variable function as such: 
\begin{equation}
    f_{h} = w_{h}(f(_{h-1})) + b_{h} \mapsto f_{h1} \circ f_{h2} \dots \circ f_{h1}
\end{equation}

The weights in each hidden layer function are like the unitary parameters in the PQC Eq. \ref{eq: pqc equation} for our sequence of gates. The network connectivity itself is modeled by the circuit ansatz with various layouts described in Sec.\ref{sec:geometry}, where hidden layers are modeled by gate operations and connectivity is modeled by entangling gates and mid-circuit measurements. Bias does not have a direct mapping for the general VQA case. \newline

\textbf{Output Function}
At each layer, an activation function is applied before being propogated into the next layer. In total, the output of a NN function is such, where $g$ is the classification task at hand:
\begin{equation}
    f_{out} =g \circ f_{h} \circ f_{h-1} \dots \circ f_{in}
\end{equation}

\textbf{Cost Function} The cost function for a NN aims to minimize the difference between predicted values and the target values and is computed as such:
\begin{equation}
    C(w,b) = \frac{1}{N}\frac{||y(x) - a_{out(x)}||}{2}
\end{equation}
In the PQC model, the cost function in Eq. \ref{eq:cost} takes on this role to compute the difference between an initial state and the target unitary that encodes the problem. Parameter updating follows the same form as in back-propagation for NN, where gradients are updated via the partial derivatives of the loss/cost function as in Eq. \ref{eq:partial}.
\subsubsection{\textbf{Universality}}

However, a key difference is in encoding the overall network. NN utilize activation functions, $a$, to model non-linearity and compute the parameter gradients such that every layer from Eq. \ref{eq:hidden-layer} is as such:
\begin{equation} \label{eq:activation}
     f_{i} = a_{i} (w_{i}*X + b_{i})
\end{equation}
Any and all PQC models as of today do not have a structured mathematical formalism for non-linearity; All PQCs take a heuristic-based approach to model the behavior of non-linearity through entangling gates. How then, can we guarantee that we will arrive at our target unitary? Well, the Solovay-Kitaev theorem in Sec. \ref{sec:unitary-group} assures us that any unitary on a large enough circuit can be approximated within an error bound of precision; it just might not be that the circuit was practical or efficient to implement, meaning the it had exponential depth~\cite{dawson_solovay-kitaev_2005}. The Solovay-Kitaev theorem is the quantum analog to the following theorem for a NN: \newline

\textbf{Universal Approximation Theorem}
For any continuous function, $f: [0,1] \mapsto [0,1]$ can be approximated arbitrarily well by a neural network with at least one hidden layer and a finite number of weights. \newline

This result can be extended by varying the upper-bound and co-domain for non-linear functions and NN of various widths to state their expressive power~\cite{gonon_universal_2023}.

\subsubsection{\textbf{The Quantum Circuit}}
Implementing the PQC circuit in Eq. \ref{eq: pqc equation}, out data going into the circuit is often prepared in the following process to encode a problem into a quantum circuit~\cite{abbas_power_2021}:
\begin{enumerate}
    \item Transform classical samples with data-prepossessing from training set, $X \in \mathcal{D}$
    \item Encode data to parameters of encoder circuit, $U_f{X}$
    \item Use variational circuit, $U(\theta)$, to derive sets of expectation values or expectation value for problem at hand.
\end{enumerate}

In this work, we focus largely on the design of the variational circuit. In practice, the circuit design follows a fixed structure of unitary gates as in Eq. \ref{eq: pqc equation}. Though The dimensionality of the vector space grows as $2^{n}$, the fixed circuit model reduces complexity by having free parameters scale polynomially with qubit count rather than exponentially~\cite{pesah_absence_2021}. Thus, our goal is to remain scalable with \textbf{parameters} and \textbf{circuit depth}. Generally circuit architecture just like connectivity in a NN is a key strategy. Below are two of the proposals to implement scalable $U(\theta)$ circuits.

\subsubsection{\textbf{NISQ Efficient Ansatz}}
Since current hardware suffers largely from noise, decoherence, and limited qubit connectivity, the hardware efficient ansatz introduced in Sec. \ref{sec:HEA} is designed to comply with the fact of sparse qubit connectivity and limited qubits. A more detailed construction of this ansatz can be found in Sec \ref{sec:HEA}. But the layouts described are generally designed to comply with connectivity and gate sets of superconducting and trapped ion computers, respectively with native entangling gates~\cite{kjaergaard2020superconducting, bruzewicz2019trapped}.

\subsubsection{\textbf{Tensor Network Ansatz}} \label{sec:TNA-ML}
The mapping from the algebraic structure of a tensor network to a quantum circuit ansatz was discussed in Sec. \ref{sec:TNA}. Classically, tensor network models were used to approximate locally and highly entangled quantum many-body systems~\cite{martin_combining_2023}. Tensor networks are methods to efficiently represent quantum states in terms of smaller interconnected tensors. In particular,these are often used to describe states whose entanglement is constrained by local interactions. By looking only at a smaller portion of the vector space, the computational cost is then reduced and becomes a polynomial function of the system size~\cite{araz2022classical}. This enables the numerical treatment of systems through layers of abstraction, reminiscent of deep neural networks. However, classic tensor networks require exponentially more data to achieve sufficient representation of the data with increasing auxiliary and Hilbert space dimensions. Thus, Quantum Tensor Networks may be significantly better than Tensor networks with a fraction of trainable parameters when the hardware is better able to simulate connective topologies beyond 1D and 2D cases. Shown in Fig. \ref{fig:enter-label} are the quantum circuit implementations for the other types of ansatz mentioned in Sec \ref{sec:TNA}.
\begin{figure*}
    \centering
    \includegraphics[scale =.5]{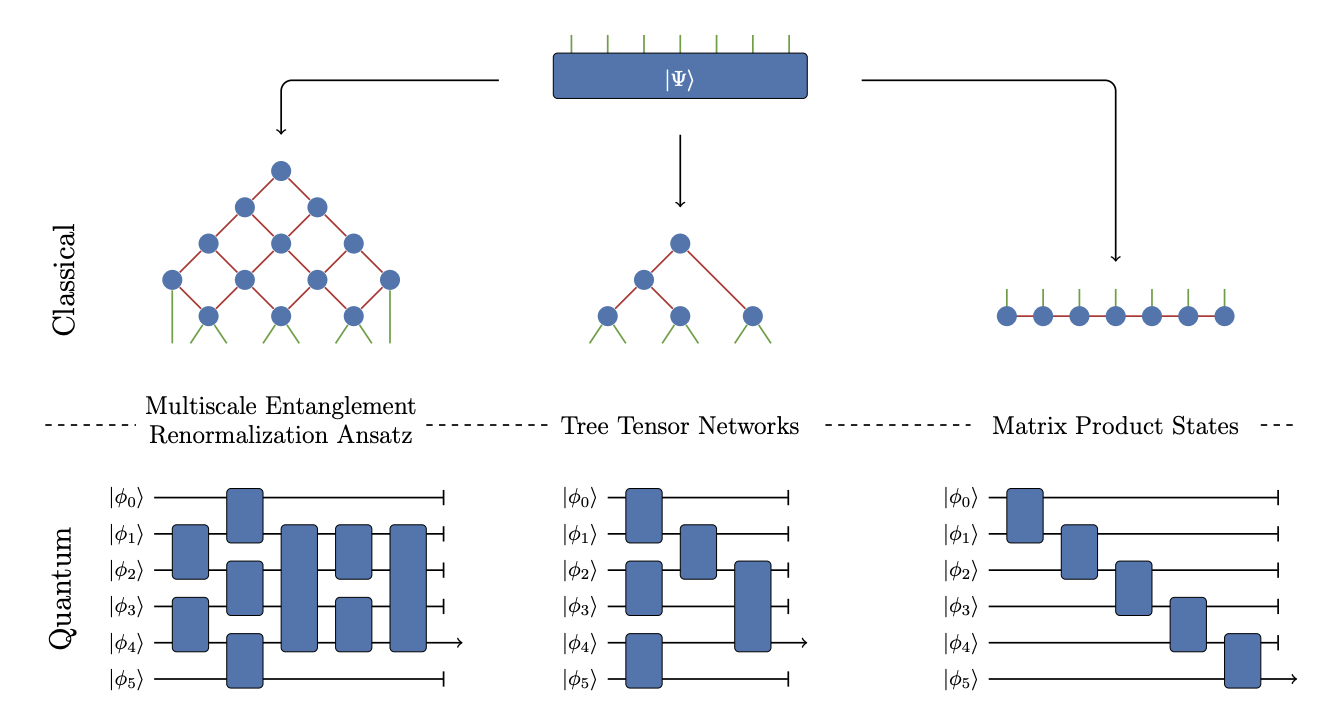}
    \caption{Different classical layouts for the MERA, TNN, and MPS ansatz and their quantum circuit implementation ~\cite{araz2022classical}}
    \label{fig:enter-label}
\end{figure*}
\subsection{Vanishing Gradient versus Barren Plateaus} \label{sec:vanishing-barren}

The analog of the barren plateau phenomenon as described in Def. \ref{sec:bp} in classical machine learning literature is the vanishing gradient for deep neural networks~\cite{tan2019vanishing}. However, the fundamental distinction for a PQC lies in the inherent impossibility of accessing the quantum state at intermediate stages during computation. While it is conceivable to employ measurements of ancillary quantum variables for the extraction of partial information, any attempt to observe the entire state of the system would inevitably perturb its quantum properties~\cite{larocca_diagnosing_2022}. Consequently, the execution of a variational circuit cannot be analogously equated with the forward pass of a neural network. Furthermore, the development of a circuit learning algorithm closely mirroring backpropagation becomes a formidable challenge due to the necessity of preserving the intermediate state of the network throughout computation. First, we will formalize the definition of backpropagation to then draw a distinction between vanishing gradients and barren plateaus~\cite{lecun1988theoretical}. \newline

 In both a neural network model and our PQC model the parameter update function for gradient-based takes on the following form:
\begin{equation} \label{eq:param-update}
    \theta(t +1) = \theta(t) - \eta\frac{\partial\mathcal{L}}{\partial\theta}
\end{equation}
For a neural network model, the partial derivative for a weight ($\theta_{jk}$ connecting the kth layer to the next jth layer) can be decomposed in the following way for a set of inputs within a dataset, 
$X\in \mathcal{D}$ we have:

\begin{equation}\label{eq:partial-derivative-ml}
    \sum_{X \in \mathcal{D}} \sum_{i_{L}=1}^{n_{L}}\sum_{i_{l}=1}^{n_{l}}\epsilon_{i_{L;X}} \frac{\partial f_{i_{L;X}}^{L}}{\partial f_{i_{l;X}}^{l}} \frac{\partial f_{i_{l;X}}^{l}}{\partial \theta_{jk}}
\end{equation}
To break apart this equation,we have an $L$-layer neural network with $n_{L}$ nodes, and we are interested in a $\theta$ at the $l$-th layer with $n_{l}$ nodes. The first term in the product can be described as:
\begin{equation}
    \frac{\partial\mathcal{L}}{\partial f_{out_{i};{X_{+}}}^{L}} = z_{out_{i}}(x_{+}; \theta) - y_{i; X_{+}}
\end{equation}

Where this term is the prediction error for the model for all its outputs versus the actual labels. There are as many $f_{out}$ equations as possible labels, so we compute the loss over the width, $n_{L}$ of the final layer in Eq. \ref{eq:partial-derivative-ml}.\newline
The middle term can be expressed as:
\begin{equation}\label{eq:partial-chain}
    \frac{\partial f_{i_{L;X}}^{L}}{\partial f_{i_{l;X}}^{l}} = \sum_{i_{l+1} \dots i_{L-1}} \prod_{l=l}^{L-1} [\theta_{jk}] a_{i_{l}}^{' l}
\end{equation}
which represents back-propagation of the errors to the $l$-th layer as a sum of products. Since every $f_{h}$ as described in Eq. \ref{eq:hidden-layer} is a composite function of an activation function with the previous $f_{l-1}$ as input,  computing the partial derivative of a path from L to l entails taking a product of $(L-1) - l$ terms of the partial derivatives until we reach the desired layer.The sum arises from the fact that each layer may have multiple paths connecting it to the next layer. \newline

The last term is known as the trivial factor, where:
\begin{equation}
     \frac{\partial f_{i_{l;X}}^{l}}{\partial b_{j}} = \delta_{ij} , \space  \frac{\partial f_{i_{l;X}}^{l}}  {\partial \theta_{jk}} =  \delta_{ij} a_{k}^{ l-1}
\end{equation}
Here, the bias term, $b$ is a parameter from Eq. \ref{eq:activation} for each layer. The bias term is a constant and there is one for each layer, so the partial derivative is just 1 when we are the corresponding layer; similarly, the derivative of Eq. \ref{eq:activation} w.r.t the weight is just the activation function from the previous layer which that weight affects.
\subsubsection{\textbf{Vanishing Gradients}}
The multiplicative factor of the partial chain in Eq. \ref{eq:partial-chain} is what gives rise to vanishing gradients through the backpropogation process in deep learning;  The gradient for a weight in a neuron depends on the sum of all the paths connecting that neuron to the output. When there are many layers, the error from activation functions are magnified, leading to large effects for earlier layers leading to a exponential decrease in the gradient. On the contrary, large prediction errors lead to exploding gradients where the magnitude of the gradient is large and thus, never converges~\cite{hanin2018neural}. fine tuning methods are needed to modulate the impact of this multiplicative factor.

\subsubsection{\textbf{Barren Plateaus}}
 Vanishing gradient arose from a dynamic process of error propagation that could be controlled with modifications the activation function or parameter initialization; barren plateaus, on the other-hand, have no means of occurring through a dynamic process such as backpropogation in a PQC, as there is no way to access intermediate information of a circuit. We have discussed various cases of barren plateaus in the quantum case in Sec. \ref{sec:bp} that were largely related to the circuit design (Sec. \ref{sec:geometry}), degree of  expressibility(Sec. \ref{sec: expressibility}), and the circuit parameterization landscape(Sec. \ref{sec:paramater-landscape}). Thus, a barren plateau can be regarded as a \textbf{static} phenomenon that appears everywhere in a circuit that is a consequence of the circuit ansatz features more closely linked to the underlying geometry.

 \subsubsection{\textbf{Variance Formalism: Quantum and Classical}}

 A barren plateau in quantum literature is defined by the average magnitudes of the cost gradients being exponentially suppressed as in Def. \ref{sec:bp} which limits the circuit  expressibility. Is there such a link in classical literature which defines the suppression of gradient updates? Looking back at the parameter update formula in Eq. \ref{eq:param-update}, the new parameter has a dependence on the magnitude of the loss function gradient, $\partial\mathcal{L}$. For the quantum case, we considered the haar integral over our unitary ensemble. From ~\cite{puchala2011symbolic}, since our haar measure is the second moment of $U(N)$, we can express it as:
 \begin{equation}
     \int_{U(N)}dUU_{ij}U_{kl}^{\dagger} = \frac{\delta_{ij}\delta_{kl}}{dim(H)} 
 \end{equation}

 Where changes in the variational angle get suppressed by the dimensionality of the Hilbert space for an expressible anastz. On the other hand, avoiding vanishing gradients requires the fine-tuning for weight initialization between layers. There is a similar relationship for the node weights initialization (with slight variation depending on the activation function~\cite{lecun1988theoretical}. between layers, the variance of weights is initialized according to the following relationship:
 \begin{equation}
     \mathbb{E}[W_{ij}W_{lk}^{\dagger}]= \frac{\sigma^{2}}{n_{l}} \delta_{ij}\delta_{lk}
 \end{equation}
 where W is the weight matrix, and the left hand side is the covariance matrix. Since all weights are assumed i.i.d., the off-diagonals are 0 and thus we get the dirac-deltas for the diagonal elements, which are the variances for each layer of weights. In this initialization scheme, the variance is scaled by the width of the layers. Thus, vanishing gradients are a consequence of large-width neural networks as the width suppress the gradient term~\cite{arora_exact_2019}.
 
\subsubsection{\textbf{Lazy Training}} \label{sec:lazy-training}
 In both the quantum and classical case, over-representing(in other words, overparameterizing) the space could potentially lead to barren plateaus or vanishing gradients where parameter updates fail~\cite{larocca_theory_2023}. Thus, the connection between large L in PQCs is analagous to the large width-case of classical neural networks. In both fields, when the value of the gradient loss stays close to its initial values, this is known as lazy training in gradient-based methods~\cite{abedi_quantum_2023}. \newline

\subsection{Neural Tangent Kernels}
In the last section we showed how quantum and classical models have suppression of the gradient updates due to large-widths. Lets now discuss a formalization of large width for classical neural networks and then the quantum case of large layers. Previously, we touched upon  expressibility of PQCs; when a PQC is overparameterized it has a remarkable ability to explore the entire state space (Sec. \ref{sec: expressibility}; however, this did lead to exponentially vanishing gradients in the context of its trainability. Classical neural networks in the case of large-width could converge linearly to zero training loss and to a global minimum with lazy training~\cite{arora_exact_2019}. What is the value from such small parameterization approximations? \newline

As it turns out, lazy training leads to a Gaussian process for the large-width limit where parameterization is replaced with a linearized model equipped with a kernel known as the Neural Tangent kernel (NKT)~\cite{jacot2018neural}. The purpose of a kernel function is to assess some notion of similarity between two feature maps, $K: \mathcal{X} \times \mathcal{X} \mapsto \mathbb{R}$. The Gaussian relation arises due to fact that output functions contain a multivariate Gaussian distribution with a covariance kernel, allowing us to compute analytical solution obtained from conditional probability~\cite{roberts_principles_2022} .Thus, the NTK informs us of how updating model parameters of one sample affects other predictions with the following form: 
\begin{equation} \label{eq:NKT}
    K_{\Theta}(x, x^{'}) = \nabla_{\Theta} f(\Theta, x) \cdot \nabla f(\Theta, x^{'})
\end{equation}

Using the NKT, we can model the loss function (Eq. \ref{eq:param-update} ) as a linear approximation as such since linear regression on a feature space is equivalent to Kernel Regression:
\begin{equation}
    f(\Theta, x) \approx  f(\Theta_{0}, x) + \nabla f(\Theta_{0}, x) \cdot (\Theta - \Theta_{0})
\end{equation}
where $\Theta$ are the set of model parameters and x is some input data. Since for lazy training as described in Sec. \ref{sec:lazy-training} the $\Theta$ is close the initial angles, the linear approximation is accurate. \newline

NKTs are vital due to the following reasons:
\begin{enumerate}
    \item The kernel is deterministic being guided by model architecture and irrelevant to weight initialization
    \item Does not change throughout the training process.
\end{enumerate}

Most critically, they provide a theoretical framework for how to characterize the dynamics of neural network training regarding their non-linear relationship transformations and how they generalize feature relationships to converge to the global minimum. 
Though currently this is possible only for the large-width limit, it will be interesting to see how to apply NKT for different model architectures~\cite{liu2020linearity}.

\subsubsection{\textbf{Quantum Neural Tangent Kernel}}
If neural network tangents can provide a way to understand the underlying mechanisms for wide neural networks in the classical case, is it possible to develop a predictable theory for PQC evolution? It terms of circuit  expressibility, barren plateaus have been a formidable issue due to circuits reaching 2-designs \ref{sec: expressibility}. As it turns out, Quantum NKTs have been recently been studied in the regime of quantum lazy training, where variational updates are small~\cite{liu_representation_2022}. As we noted earlier, For quantum circuits, width was found to be dependent on the number of hermitian operators, where correlations are suppressed by the dimensionality of the Hilbert space. The parameters are also chosen at random, so the eigenvalues of the average tangent kernel determine the training behaviour of such parameterized quantum circuits. Interestingly, when the model is full rank, the training converges at exponential speed in the linear model--opposite of what we desire~\cite{nakaji_quantum-enhanced_2023}. However, to mitigate the barren plateau issue with exponential convergence, the circuit geometry as well as local observables must be considered instead to have a circuit enter the lazy-training phase~\cite{liu_laziness_2022}. As an extension to training speed, what role might phase transitions play in identifying critical points for the parameter landscape? Another interesting thing to note is how hardware noise will affect quantum circuits entering the lazy-training phase as well as the behavior of the quantum analog of the acquired error, and whether it is more advantageous to use a subspace for a PQC with less expressibility for training. Thus QNKT would provide a framework for understanding how designs of PQCs will be a trade-off between barren plateaus and performance.

\section{Conclusion}

In this paper, we presented the groundwork for various formalisms of barren plateaus present as roadblocks in developing PQCs efficiently. In Sec. \ref{section:background}, we established the formal definitions underlying unitary distributions and the expressibility of unique ensembles related to the haar measure (Sec. \ref{sec:haar}). Notably, we remarked how truly random circuits are an unsuitable for hybrid quantum-classical algorithms as an initial guess to run on more than a few qubits due to the cost involved in assessing an exponentially increasing Hilbert space~\cite{bouland2019complexity}. The characterization of a circuit as a 2-design is the standard to assess the  expressibility of a unitary ensemble(Eq. \ref{eq:2-design}). Whether or not a circuit forms an approximate t-design is an integral step in evaluating the presence of a barren plateau, and we have formalized commonly used  metrics such as frame potentials(Sec. \ref{sec: frame-potential}) and state fidelity(Eq. \ref{equation: fidelity} )to assess the proximity of an ensemble to the haar distribution~\cite{harrow_random_2009}. Thus, Barren plateaus (Sec. \ref{sec:bp}) can be characterized in a probabilistic sense as an onset of exponentially vanishing gradients where the loss function expectation concentrates around 0.\newline

Next, we establish commonly used circuit ansatz schemes in developing PQCs. Most NIQS research for generalizable training and for mitigating noise opts to use the brick-work ansatz that is hardware efficient(Sec. \ref{sec:HEA}. Physically motivated ansatz are commonly used when the approximate solution space is known and thus, it is advantageous to take advantage of local symmetries within the system of interest for faster training convergence~\cite{meyer_exploiting_2023}. Another interesting ansatz is the tensor network ansatz (Sec. \ref{sec:TNA})represented by a matrix product state. In these ansatz, the goal is to take advantage of underlying local connectivity to constrain the  expressibility of the system in one part of a biparttie state through evaluating the entropy of a split. As of now only 1D and 2D connectivity models have been studied, so there is room for researching how well MPS states generalize to higher dimensions. \newline

Next, we laid the foundation for what defines PQC performance through the cost-function landscape(Sec. \ref{sec:paramater-landscape}), circuit  expressibility (Sec. \ref{sec: expressibility}), and the underlying circuit geometry(Sec. \ref{sec:geometry}). Each of these components themselves have metrics to quantify the impact on trainability.  expressibility is itself a trade-off between how much a circuit is able to explore the unitary space and the likelihood of avoiding a barren plateau. As it turns out, barren plateaus may have characterizations through the underlying circuit geometry  that may lead to is better techniques aside from gradient descent, most notably using the Quantum fisher information matrix to assess the eigenvalues of a state-space~\cite{liu_quantum_2020}.  QFIM is used in quantum estimation theory to quantify the sensitivity of a quantum state to changes in
parameters and characterizes how the state’s probability distribution changes with variations in parameters; for circuit geometry, the underlying landscape as other topological structures such as fiber bundles is a better fit for such metrics that can take into account complex volumetric relationships and physical transformations~\cite{haegeman_geometry_2014}. \newline

We provided various factors underlying barren plateaus in current literature. Most notable whether a cost function (Sec. \ref{sec:cost-bp} is global or local as well as the depth of a circuit is strongly tied to the formation of a barren plateau~\cite{cerezo2021cost}. Thus, using shallow depth circuits that are expressive enough is vital, leading to the exploration of parameter initialization strategies. Noise-induced barren plateaus are notoriously difficult to control, as they propagate hardware gate errors and lead to a deterministic outcome for barren plateaus in the limit of circuit depth~\cite{wang_noise-induced_2021}. (Sec. \ref{sec:noise-bp}). There are not many remedies other than choice of ansatz that best aligns with the underlying hardware at the moment. The final induced barren plateau mentioned was through entanglement. (Sec. \ref{sec: entanglement}).Here, we characterized the volume and area law growth of entanglement in a system and how that creates information scrambling, leading to an inability to locate the target ground state in polynomial time~\cite{marrero_entanglement_2021}. Assessing a sort of "pre-image" of a barren plateau to adaptively reset the learning rate or re-initialize parameters is a strategy to modulate a circuit reaching its entropy saturation point.\newline

Finally, we concluded with the relationships between barren plateaus and vanishing gradients in classical machine learning (Sec. \ref{sec:vanishing-barren})~\cite{sharma_trainability_2022}. We characterized how a PQC draws inspiration from a neural network model an the underlying classical challenges that lead to vanishing gradients. Inherently, vanishing gradients are different due to the way they manifest through backpropagation for the loss function and are an analytical process-driven consequence, whereas barren plateaus are tied more closely with the underlying geometry of a circuit.  We made a remark on the utility of using kernel functions to characterize the behavior of trainability in the limit of large operator circuits.  \newline

In all, the future of PQCs in the NISQ area is most prominent in establishing problems closely modeling by physics-inspired behavior and where circuit designs make take advantage of system and Hamiltonian symmetries to best mitigate the onset of barren plateaus. It will also be interesting how much machine-learning theory for neural networks can guide parameterization strategies for variational circuits in the future.

\bibliographystyle{apsrev4-1} 
\bibliography{citations} 

\begin{thebibliography}{101}%
\makeatletter
\providecommand \@ifxundefined [1]{%
 \@ifx{#1\undefined}
}%
\providecommand \@ifnum [1]{%
 \ifnum #1\expandafter \@firstoftwo
 \else \expandafter \@secondoftwo
 \fi
}%
\providecommand \@ifx [1]{%
 \ifx #1\expandafter \@firstoftwo
 \else \expandafter \@secondoftwo
 \fi
}%
\providecommand \natexlab [1]{#1}%
\providecommand \enquote  [1]{``#1''}%
\providecommand \bibnamefont  [1]{#1}%
\providecommand \bibfnamefont [1]{#1}%
\providecommand \citenamefont [1]{#1}%
\providecommand \href@noop [0]{\@secondoftwo}%
\providecommand \href [0]{\begingroup \@sanitize@url \@href}%
\providecommand \@href[1]{\@@startlink{#1}\@@href}%
\providecommand \@@href[1]{\endgroup#1\@@endlink}%
\providecommand \@sanitize@url [0]{\catcode `\\12\catcode `\$12\catcode `\&12\catcode `\#12\catcode `\^12\catcode `\_12\catcode `\%12\relax}%
\providecommand \@@startlink[1]{}%
\providecommand \@@endlink[0]{}%
\providecommand \url  [0]{\begingroup\@sanitize@url \@url }%
\providecommand \@url [1]{\endgroup\@href {#1}{\urlprefix }}%
\providecommand \urlprefix  [0]{URL }%
\providecommand \Eprint [0]{\href }%
\providecommand \doibase [0]{http://dx.doi.org/}%
\providecommand \selectlanguage [0]{\@gobble}%
\providecommand \bibinfo  [0]{\@secondoftwo}%
\providecommand \bibfield  [0]{\@secondoftwo}%
\providecommand \translation [1]{[#1]}%
\providecommand \BibitemOpen [0]{}%
\providecommand \bibitemStop [0]{}%
\providecommand \bibitemNoStop [0]{.\EOS\space}%
\providecommand \EOS [0]{\spacefactor3000\relax}%
\providecommand \BibitemShut  [1]{\csname bibitem#1\endcsname}%
\let\auto@bib@innerbib\@empty
\bibitem [{\citenamefont {Du}\ \emph {et~al.}({\natexlab{a}})\citenamefont {Du}, \citenamefont {Huang}, \citenamefont {You}, \citenamefont {Hsieh},\ and\ \citenamefont {Tao}}]{du_quantum_2022}%
  \BibitemOpen
  \bibfield  {author} {\bibinfo {author} {\bibfnamefont {Y.}~\bibnamefont {Du}}, \bibinfo {author} {\bibfnamefont {T.}~\bibnamefont {Huang}}, \bibinfo {author} {\bibfnamefont {S.}~\bibnamefont {You}}, \bibinfo {author} {\bibfnamefont {M.-H.}\ \bibnamefont {Hsieh}}, \ and\ \bibinfo {author} {\bibfnamefont {D.}~\bibnamefont {Tao}},\ }\href {\doibase 10.1038/s41534-022-00570-y} {\ \textbf {\bibinfo {volume} {8}},\ \bibinfo {pages} {62} ({\natexlab{a}})},\ \Eprint {http://arxiv.org/abs/2010.10217 [quant-ph]} {2010.10217 [quant-ph]} \BibitemShut {NoStop}%
\bibitem [{\citenamefont {Leone}\ \emph {et~al.}()\citenamefont {Leone}, \citenamefont {Oliviero}, \citenamefont {Cincio},\ and\ \citenamefont {Cerezo}}]{leone_practical_2022}%
  \BibitemOpen
  \bibfield  {author} {\bibinfo {author} {\bibfnamefont {L.}~\bibnamefont {Leone}}, \bibinfo {author} {\bibfnamefont {S.~F.~E.}\ \bibnamefont {Oliviero}}, \bibinfo {author} {\bibfnamefont {L.}~\bibnamefont {Cincio}}, \ and\ \bibinfo {author} {\bibfnamefont {M.}~\bibnamefont {Cerezo}},\ }\href {http://arxiv.org/abs/2211.01477} {\enquote {\bibinfo {title} {On the practical usefulness of the hardware efficient ansatz},}\ }\Eprint {http://arxiv.org/abs/2211.01477 [quant-ph]} {2211.01477 [quant-ph]} \BibitemShut {NoStop}%
\bibitem [{\citenamefont {Cao}\ \emph {et~al.}()\citenamefont {Cao}, \citenamefont {Romero}, \citenamefont {Olson}, \citenamefont {Degroote}, \citenamefont {Johnson}, \citenamefont {Kieferová}, \citenamefont {Kivlichan}, \citenamefont {Menke}, \citenamefont {Peropadre}, \citenamefont {Sawaya}, \citenamefont {Sim}, \citenamefont {Veis},\ and\ \citenamefont {Aspuru-Guzik}}]{cao_quantum_2019}%
  \BibitemOpen
  \bibfield  {author} {\bibinfo {author} {\bibfnamefont {Y.}~\bibnamefont {Cao}}, \bibinfo {author} {\bibfnamefont {J.}~\bibnamefont {Romero}}, \bibinfo {author} {\bibfnamefont {J.~P.}\ \bibnamefont {Olson}}, \bibinfo {author} {\bibfnamefont {M.}~\bibnamefont {Degroote}}, \bibinfo {author} {\bibfnamefont {P.~D.}\ \bibnamefont {Johnson}}, \bibinfo {author} {\bibfnamefont {M.}~\bibnamefont {Kieferová}}, \bibinfo {author} {\bibfnamefont {I.~D.}\ \bibnamefont {Kivlichan}}, \bibinfo {author} {\bibfnamefont {T.}~\bibnamefont {Menke}}, \bibinfo {author} {\bibfnamefont {B.}~\bibnamefont {Peropadre}}, \bibinfo {author} {\bibfnamefont {N.~P.~D.}\ \bibnamefont {Sawaya}}, \bibinfo {author} {\bibfnamefont {S.}~\bibnamefont {Sim}}, \bibinfo {author} {\bibfnamefont {L.}~\bibnamefont {Veis}}, \ and\ \bibinfo {author} {\bibfnamefont {A.}~\bibnamefont {Aspuru-Guzik}},\ }\href {\doibase 10.1021/acs.chemrev.8b00803} {\ \textbf {\bibinfo {volume} {119}},\ \bibinfo {pages} {10856}},\ \Eprint {http://arxiv.org/abs/1812.09976
  [quant-ph]} {1812.09976 [quant-ph]} \BibitemShut {NoStop}%
\bibitem [{noa({\natexlab{a}})}]{noauthor_practical_nodate}%
  \BibitemOpen
  \href {https://www.nature.com/articles/s41586-022-04940-6} {\enquote {\bibinfo {title} {Practical quantum advantage in quantum simulation {\textbar} nature},}\ } ({\natexlab{a}})\BibitemShut {NoStop}%
\bibitem [{\citenamefont {Shor}()}]{shor_fault-tolerant_1997}%
  \BibitemOpen
  \bibfield  {author} {\bibinfo {author} {\bibfnamefont {P.~W.}\ \bibnamefont {Shor}},\ }\href {http://arxiv.org/abs/quant-ph/9605011} {\enquote {\bibinfo {title} {Fault-tolerant quantum computation},}\ }\Eprint {http://arxiv.org/abs/quant-ph/9605011} {quant-ph/9605011} \BibitemShut {NoStop}%
\bibitem [{\citenamefont {Bottou}()}]{bottou_large-scale_2010}%
  \BibitemOpen
  \bibfield  {author} {\bibinfo {author} {\bibfnamefont {L.}~\bibnamefont {Bottou}},\ }in\ \href {\doibase 10.1007/978-3-7908-2604-3_16} {\emph {\bibinfo {booktitle} {Proceedings of {COMPSTAT}'2010}}},\ \bibinfo {editor} {edited by\ \bibinfo {editor} {\bibfnamefont {Y.}~\bibnamefont {Lechevallier}}\ and\ \bibinfo {editor} {\bibfnamefont {G.}~\bibnamefont {Saporta}}}\ (\bibinfo  {publisher} {Physica-Verlag {HD}})\ pp.\ \bibinfo {pages} {177--186}\BibitemShut {NoStop}%
\bibitem [{\citenamefont {LeCun}\ \emph {et~al.}(2002)\citenamefont {LeCun}, \citenamefont {Bottou}, \citenamefont {Orr},\ and\ \citenamefont {M{\"u}ller}}]{lecun2002efficient}%
  \BibitemOpen
  \bibfield  {author} {\bibinfo {author} {\bibfnamefont {Y.}~\bibnamefont {LeCun}}, \bibinfo {author} {\bibfnamefont {L.}~\bibnamefont {Bottou}}, \bibinfo {author} {\bibfnamefont {G.~B.}\ \bibnamefont {Orr}}, \ and\ \bibinfo {author} {\bibfnamefont {K.-R.}\ \bibnamefont {M{\"u}ller}},\ }in\ \href@noop {} {\emph {\bibinfo {booktitle} {Neural networks: Tricks of the trade}}}\ (\bibinfo  {publisher} {Springer},\ \bibinfo {year} {2002})\ pp.\ \bibinfo {pages} {9--50}\BibitemShut {NoStop}%
\bibitem [{\citenamefont {Abbas}\ \emph {et~al.}()\citenamefont {Abbas}, \citenamefont {Sutter}, \citenamefont {Zoufal}, \citenamefont {Lucchi}, \citenamefont {Figalli},\ and\ \citenamefont {Woerner}}]{abbas_power_2021}%
  \BibitemOpen
  \bibfield  {author} {\bibinfo {author} {\bibfnamefont {A.}~\bibnamefont {Abbas}}, \bibinfo {author} {\bibfnamefont {D.}~\bibnamefont {Sutter}}, \bibinfo {author} {\bibfnamefont {C.}~\bibnamefont {Zoufal}}, \bibinfo {author} {\bibfnamefont {A.}~\bibnamefont {Lucchi}}, \bibinfo {author} {\bibfnamefont {A.}~\bibnamefont {Figalli}}, \ and\ \bibinfo {author} {\bibfnamefont {S.}~\bibnamefont {Woerner}},\ }\href {\doibase 10.1038/s43588-021-00084-1} {\ \textbf {\bibinfo {volume} {1}},\ \bibinfo {pages} {403}},\ \bibinfo {note} {number: 6 Publisher: Nature Publishing Group}\BibitemShut {NoStop}%
\bibitem [{\citenamefont {De~Palma}\ \emph {et~al.}()\citenamefont {De~Palma}, \citenamefont {Marvian}, \citenamefont {Rouzé},\ and\ \citenamefont {França}}]{de_palma_limitations_2023}%
  \BibitemOpen
  \bibfield  {author} {\bibinfo {author} {\bibfnamefont {G.}~\bibnamefont {De~Palma}}, \bibinfo {author} {\bibfnamefont {M.}~\bibnamefont {Marvian}}, \bibinfo {author} {\bibfnamefont {C.}~\bibnamefont {Rouzé}}, \ and\ \bibinfo {author} {\bibfnamefont {D.~S.}\ \bibnamefont {França}},\ }\href {\doibase 10.1103/PRXQuantum.4.010309} {\ \textbf {\bibinfo {volume} {4}},\ \bibinfo {pages} {010309}},\ \bibinfo {note} {publisher: American Physical Society}\BibitemShut {NoStop}%
\bibitem [{\citenamefont {Wiersema}\ \emph {et~al.}({\natexlab{a}})\citenamefont {Wiersema}, \citenamefont {Lewis}, \citenamefont {Wierichs}, \citenamefont {Carrasquilla},\ and\ \citenamefont {Killoran}}]{wiersema_here_2023}%
  \BibitemOpen
  \bibfield  {author} {\bibinfo {author} {\bibfnamefont {R.}~\bibnamefont {Wiersema}}, \bibinfo {author} {\bibfnamefont {D.}~\bibnamefont {Lewis}}, \bibinfo {author} {\bibfnamefont {D.}~\bibnamefont {Wierichs}}, \bibinfo {author} {\bibfnamefont {J.}~\bibnamefont {Carrasquilla}}, \ and\ \bibinfo {author} {\bibfnamefont {N.}~\bibnamefont {Killoran}},\ }\href {http://arxiv.org/abs/2303.11355} {\enquote {\bibinfo {title} {Here comes the \${\textbackslash}mathrm\{{SU}\}(n)\$: multivariate quantum gates and gradients},}\ } ({\natexlab{a}}),\ \Eprint {http://arxiv.org/abs/2303.11355 [quant-ph]} {2303.11355 [quant-ph]} \BibitemShut {NoStop}%
\bibitem [{\citenamefont {Harrow}\ and\ \citenamefont {Low}()}]{harrow_random_2009}%
  \BibitemOpen
  \bibfield  {author} {\bibinfo {author} {\bibfnamefont {A.~W.}\ \bibnamefont {Harrow}}\ and\ \bibinfo {author} {\bibfnamefont {R.~A.}\ \bibnamefont {Low}},\ }\href {\doibase 10.1007/s00220-009-0873-6} {\ \textbf {\bibinfo {volume} {291}},\ \bibinfo {pages} {257}}\BibitemShut {NoStop}%
\bibitem [{\citenamefont {Hunter-Jones}({\natexlab{a}})}]{hunter-jones_unitary_2019}%
  \BibitemOpen
  \bibfield  {author} {\bibinfo {author} {\bibfnamefont {N.}~\bibnamefont {Hunter-Jones}},\ }\href {http://arxiv.org/abs/1905.12053} {\enquote {\bibinfo {title} {Unitary designs from statistical mechanics in random quantum circuits},}\ } ({\natexlab{a}}),\ \Eprint {http://arxiv.org/abs/1905.12053 [cond-mat, physics:hep-th, physics:quant-ph]} {1905.12053 [cond-mat, physics:hep-th, physics:quant-ph]} \BibitemShut {NoStop}%
\bibitem [{\citenamefont {Wang}\ \emph {et~al.}()\citenamefont {Wang}, \citenamefont {Fontana}, \citenamefont {Cerezo}, \citenamefont {Sharma}, \citenamefont {Sone}, \citenamefont {Cincio},\ and\ \citenamefont {Coles}}]{wang_noise-induced_2021}%
  \BibitemOpen
  \bibfield  {author} {\bibinfo {author} {\bibfnamefont {S.}~\bibnamefont {Wang}}, \bibinfo {author} {\bibfnamefont {E.}~\bibnamefont {Fontana}}, \bibinfo {author} {\bibfnamefont {M.}~\bibnamefont {Cerezo}}, \bibinfo {author} {\bibfnamefont {K.}~\bibnamefont {Sharma}}, \bibinfo {author} {\bibfnamefont {A.}~\bibnamefont {Sone}}, \bibinfo {author} {\bibfnamefont {L.}~\bibnamefont {Cincio}}, \ and\ \bibinfo {author} {\bibfnamefont {P.~J.}\ \bibnamefont {Coles}},\ }\href {\doibase 10.1038/s41467-021-27045-6} {\ \textbf {\bibinfo {volume} {12}},\ \bibinfo {pages} {6961}},\ \Eprint {http://arxiv.org/abs/2007.14384 [quant-ph]} {2007.14384 [quant-ph]} \BibitemShut {NoStop}%
\bibitem [{\citenamefont {Patti}\ \emph {et~al.}()\citenamefont {Patti}, \citenamefont {Najafi}, \citenamefont {Gao},\ and\ \citenamefont {Yelin}}]{patti_entanglement_2021}%
  \BibitemOpen
  \bibfield  {author} {\bibinfo {author} {\bibfnamefont {T.~L.}\ \bibnamefont {Patti}}, \bibinfo {author} {\bibfnamefont {K.}~\bibnamefont {Najafi}}, \bibinfo {author} {\bibfnamefont {X.}~\bibnamefont {Gao}}, \ and\ \bibinfo {author} {\bibfnamefont {S.~F.}\ \bibnamefont {Yelin}},\ }\href {\doibase 10.1103/PhysRevResearch.3.033090} {\ \textbf {\bibinfo {volume} {3}},\ \bibinfo {pages} {033090}},\ \Eprint {http://arxiv.org/abs/2012.12658 [quant-ph]} {2012.12658 [quant-ph]} \BibitemShut {NoStop}%
\bibitem [{\citenamefont {Uvarov}\ and\ \citenamefont {Biamonte}()}]{uvarov_barren_2021}%
  \BibitemOpen
  \bibfield  {author} {\bibinfo {author} {\bibfnamefont {A.~V.}\ \bibnamefont {Uvarov}}\ and\ \bibinfo {author} {\bibfnamefont {J.~D.}\ \bibnamefont {Biamonte}},\ }\href {\doibase 10.1088/1751-8121/abfac7} {\ \textbf {\bibinfo {volume} {54}},\ \bibinfo {pages} {245301}},\ \bibinfo {note} {publisher: {IOP} Publishing}\BibitemShut {NoStop}%
\bibitem [{\citenamefont {Napp}()}]{napp_quantifying_2022}%
  \BibitemOpen
  \bibfield  {author} {\bibinfo {author} {\bibfnamefont {J.}~\bibnamefont {Napp}},\ }\href {http://arxiv.org/abs/2203.06174} {\enquote {\bibinfo {title} {Quantifying the barren plateau phenomenon for a model of unstructured variational ans{\textbackslash}"\{a\}tze},}\ }\Eprint {http://arxiv.org/abs/2203.06174 [quant-ph]} {2203.06174 [quant-ph]} \BibitemShut {NoStop}%
\bibitem [{\citenamefont {Zhu}\ \emph {et~al.}()\citenamefont {Zhu}, \citenamefont {Linke}, \citenamefont {Benedetti}, \citenamefont {Landsman}, \citenamefont {Nguyen}, \citenamefont {Alderete}, \citenamefont {Perdomo-Ortiz}, \citenamefont {Korda}, \citenamefont {Garfoot}, \citenamefont {Brecque}, \citenamefont {Egan}, \citenamefont {Perdomo},\ and\ \citenamefont {Monroe}}]{zhu_training_2019}%
  \BibitemOpen
  \bibfield  {author} {\bibinfo {author} {\bibfnamefont {D.}~\bibnamefont {Zhu}}, \bibinfo {author} {\bibfnamefont {N.~M.}\ \bibnamefont {Linke}}, \bibinfo {author} {\bibfnamefont {M.}~\bibnamefont {Benedetti}}, \bibinfo {author} {\bibfnamefont {K.~A.}\ \bibnamefont {Landsman}}, \bibinfo {author} {\bibfnamefont {N.~H.}\ \bibnamefont {Nguyen}}, \bibinfo {author} {\bibfnamefont {C.~H.}\ \bibnamefont {Alderete}}, \bibinfo {author} {\bibfnamefont {A.}~\bibnamefont {Perdomo-Ortiz}}, \bibinfo {author} {\bibfnamefont {N.}~\bibnamefont {Korda}}, \bibinfo {author} {\bibfnamefont {A.}~\bibnamefont {Garfoot}}, \bibinfo {author} {\bibfnamefont {C.}~\bibnamefont {Brecque}}, \bibinfo {author} {\bibfnamefont {L.}~\bibnamefont {Egan}}, \bibinfo {author} {\bibfnamefont {O.}~\bibnamefont {Perdomo}}, \ and\ \bibinfo {author} {\bibfnamefont {C.}~\bibnamefont {Monroe}},\ }\href {\doibase 10.1126/sciadv.aaw9918} {\ \textbf {\bibinfo {volume} {5}},\ \bibinfo {pages} {eaaw9918}},\ \bibinfo {note} {publisher: American Association for
  the Advancement of Science}\BibitemShut {NoStop}%
\bibitem [{\citenamefont {Wecker}\ \emph {et~al.}()\citenamefont {Wecker}, \citenamefont {Hastings},\ and\ \citenamefont {Troyer}}]{wecker_towards_2015}%
  \BibitemOpen
  \bibfield  {author} {\bibinfo {author} {\bibfnamefont {D.}~\bibnamefont {Wecker}}, \bibinfo {author} {\bibfnamefont {M.~B.}\ \bibnamefont {Hastings}}, \ and\ \bibinfo {author} {\bibfnamefont {M.}~\bibnamefont {Troyer}},\ }\href {\doibase 10.1103/PhysRevA.92.042303} {\ \textbf {\bibinfo {volume} {92}},\ \bibinfo {pages} {042303}},\ \Eprint {http://arxiv.org/abs/1507.08969 [cond-mat, physics:quant-ph]} {1507.08969 [cond-mat, physics:quant-ph]} \BibitemShut {NoStop}%
\bibitem [{\citenamefont {Cerezo}\ \emph {et~al.}()\citenamefont {Cerezo}, \citenamefont {Arrasmith}, \citenamefont {Babbush}, \citenamefont {Benjamin}, \citenamefont {Endo}, \citenamefont {Fujii}, \citenamefont {{McClean}}, \citenamefont {Mitarai}, \citenamefont {Yuan}, \citenamefont {Cincio},\ and\ \citenamefont {Coles}}]{cerezo_variational_2021}%
  \BibitemOpen
  \bibfield  {author} {\bibinfo {author} {\bibfnamefont {M.}~\bibnamefont {Cerezo}}, \bibinfo {author} {\bibfnamefont {A.}~\bibnamefont {Arrasmith}}, \bibinfo {author} {\bibfnamefont {R.}~\bibnamefont {Babbush}}, \bibinfo {author} {\bibfnamefont {S.~C.}\ \bibnamefont {Benjamin}}, \bibinfo {author} {\bibfnamefont {S.}~\bibnamefont {Endo}}, \bibinfo {author} {\bibfnamefont {K.}~\bibnamefont {Fujii}}, \bibinfo {author} {\bibfnamefont {J.~R.}\ \bibnamefont {{McClean}}}, \bibinfo {author} {\bibfnamefont {K.}~\bibnamefont {Mitarai}}, \bibinfo {author} {\bibfnamefont {X.}~\bibnamefont {Yuan}}, \bibinfo {author} {\bibfnamefont {L.}~\bibnamefont {Cincio}}, \ and\ \bibinfo {author} {\bibfnamefont {P.~J.}\ \bibnamefont {Coles}},\ }\href {\doibase 10.1038/s42254-021-00348-9} {\ \textbf {\bibinfo {volume} {3}},\ \bibinfo {pages} {625}},\ \bibinfo {note} {number: 9 Publisher: Nature Publishing Group}\BibitemShut {NoStop}%
\bibitem [{\citenamefont {Nakaji}\ and\ \citenamefont {Yamamoto}()}]{nakaji_expressibility_2021}%
  \BibitemOpen
  \bibfield  {author} {\bibinfo {author} {\bibfnamefont {K.}~\bibnamefont {Nakaji}}\ and\ \bibinfo {author} {\bibfnamefont {N.}~\bibnamefont {Yamamoto}},\ }\href {\doibase 10.22331/q-2021-04-19-434} {\ \textbf {\bibinfo {volume} {5}},\ \bibinfo {pages} {434}},\ \Eprint {http://arxiv.org/abs/2005.12537 [quant-ph]} {2005.12537 [quant-ph]} \BibitemShut {NoStop}%
\bibitem [{\citenamefont {{ITZYKSON}}\ and\ \citenamefont {{NAUENBERG}}()}]{itzykson_unitary_1966}%
  \BibitemOpen
  \bibfield  {author} {\bibinfo {author} {\bibfnamefont {C.}~\bibnamefont {{ITZYKSON}}}\ and\ \bibinfo {author} {\bibfnamefont {M.}~\bibnamefont {{NAUENBERG}}},\ }\href {\doibase 10.1103/RevModPhys.38.95} {\ \textbf {\bibinfo {volume} {38}},\ \bibinfo {pages} {95}},\ \bibinfo {note} {publisher: American Physical Society}\BibitemShut {NoStop}%
\bibitem [{\citenamefont {Isaev}\ and\ \citenamefont {Kruzhilin}(2002)}]{isaev2002effective}%
  \BibitemOpen
  \bibfield  {author} {\bibinfo {author} {\bibfnamefont {A.}~\bibnamefont {Isaev}}\ and\ \bibinfo {author} {\bibfnamefont {N.}~\bibnamefont {Kruzhilin}},\ }\href@noop {} {\bibfield  {journal} {\bibinfo  {journal} {Canadian Journal of Mathematics}\ }\textbf {\bibinfo {volume} {54}},\ \bibinfo {pages} {1254} (\bibinfo {year} {2002})}\BibitemShut {NoStop}%
\bibitem [{\citenamefont {Gamburd}\ \emph {et~al.}(1999)\citenamefont {Gamburd}, \citenamefont {Sarnak},\ and\ \citenamefont {Jakobson}}]{gamburd1999spectra}%
  \BibitemOpen
  \bibfield  {author} {\bibinfo {author} {\bibfnamefont {A.}~\bibnamefont {Gamburd}}, \bibinfo {author} {\bibfnamefont {P.}~\bibnamefont {Sarnak}}, \ and\ \bibinfo {author} {\bibfnamefont {D.}~\bibnamefont {Jakobson}},\ }\href@noop {} {\bibfield  {journal} {\bibinfo  {journal} {Journal of the European Mathematical Society}\ }\textbf {\bibinfo {volume} {1}},\ \bibinfo {pages} {51} (\bibinfo {year} {1999})}\BibitemShut {NoStop}%
\bibitem [{\citenamefont {Dawson}\ and\ \citenamefont {Nielsen}()}]{dawson_solovay-kitaev_2005}%
  \BibitemOpen
  \bibfield  {author} {\bibinfo {author} {\bibfnamefont {C.~M.}\ \bibnamefont {Dawson}}\ and\ \bibinfo {author} {\bibfnamefont {M.~A.}\ \bibnamefont {Nielsen}},\ }\href {http://arxiv.org/abs/quant-ph/0505030} {\enquote {\bibinfo {title} {The solovay-kitaev algorithm},}\ }\Eprint {http://arxiv.org/abs/quant-ph/0505030} {quant-ph/0505030} \BibitemShut {NoStop}%
\bibitem [{\citenamefont {Meyer}\ \emph {et~al.}()\citenamefont {Meyer}, \citenamefont {Mularski}, \citenamefont {Gil-Fuster}, \citenamefont {Mele}, \citenamefont {Arzani}, \citenamefont {Wilms},\ and\ \citenamefont {Eisert}}]{meyer_exploiting_2023}%
  \BibitemOpen
  \bibfield  {author} {\bibinfo {author} {\bibfnamefont {J.~J.}\ \bibnamefont {Meyer}}, \bibinfo {author} {\bibfnamefont {M.}~\bibnamefont {Mularski}}, \bibinfo {author} {\bibfnamefont {E.}~\bibnamefont {Gil-Fuster}}, \bibinfo {author} {\bibfnamefont {A.~A.}\ \bibnamefont {Mele}}, \bibinfo {author} {\bibfnamefont {F.}~\bibnamefont {Arzani}}, \bibinfo {author} {\bibfnamefont {A.}~\bibnamefont {Wilms}}, \ and\ \bibinfo {author} {\bibfnamefont {J.}~\bibnamefont {Eisert}},\ }\href {\doibase 10.1103/PRXQuantum.4.010328} {\ \textbf {\bibinfo {volume} {4}},\ \bibinfo {pages} {010328}},\ \Eprint {http://arxiv.org/abs/2205.06217 [quant-ph]} {2205.06217 [quant-ph]} \BibitemShut {NoStop}%
\bibitem [{\citenamefont {Riser}\ and\ \citenamefont {Kanzieper}()}]{riser_power_2023}%
  \BibitemOpen
  \bibfield  {author} {\bibinfo {author} {\bibfnamefont {R.}~\bibnamefont {Riser}}\ and\ \bibinfo {author} {\bibfnamefont {E.}~\bibnamefont {Kanzieper}},\ }\href {\doibase 10.1016/j.physd.2022.133599} {\ \textbf {\bibinfo {volume} {444}},\ \bibinfo {pages} {133599}},\ \Eprint {http://arxiv.org/abs/2209.04723 [cond-mat, physics:math-ph, physics:nlin, physics:quant-ph]} {2209.04723 [cond-mat, physics:math-ph, physics:nlin, physics:quant-ph]} \BibitemShut {NoStop}%
\bibitem [{\citenamefont {Mele}()}]{mele_introduction_2023}%
  \BibitemOpen
  \bibfield  {author} {\bibinfo {author} {\bibfnamefont {A.~A.}\ \bibnamefont {Mele}},\ }\href {http://arxiv.org/abs/2307.08956} {\enquote {\bibinfo {title} {Introduction to haar measure tools in quantum information: A beginner's tutorial},}\ }\Eprint {http://arxiv.org/abs/2307.08956 [quant-ph]} {2307.08956 [quant-ph]} \BibitemShut {NoStop}%
\bibitem [{\citenamefont {Hunter-Jones}({\natexlab{b}})}]{hunter-jones_operator_2018}%
  \BibitemOpen
  \bibfield  {author} {\bibinfo {author} {\bibfnamefont {N.}~\bibnamefont {Hunter-Jones}},\ }\href {http://arxiv.org/abs/1812.08219} {\enquote {\bibinfo {title} {Operator growth in random quantum circuits with symmetry},}\ } ({\natexlab{b}}),\ \Eprint {http://arxiv.org/abs/1812.08219 [cond-mat, physics:hep-th, physics:quant-ph]} {1812.08219 [cond-mat, physics:hep-th, physics:quant-ph]} \BibitemShut {NoStop}%
\bibitem [{\citenamefont {Dakovic}\ \emph {et~al.}()\citenamefont {Dakovic}, \citenamefont {Denker},\ and\ \citenamefont {Gordin}}]{dakovic_circular_2016}%
  \BibitemOpen
  \bibfield  {author} {\bibinfo {author} {\bibfnamefont {R.}~\bibnamefont {Dakovic}}, \bibinfo {author} {\bibfnamefont {M.}~\bibnamefont {Denker}}, \ and\ \bibinfo {author} {\bibfnamefont {M.}~\bibnamefont {Gordin}},\ }\href {\doibase 10.1007/s10958-016-3141-2} {\ \textbf {\bibinfo {volume} {219}},\ \bibinfo {pages} {714}}\BibitemShut {NoStop}%
\bibitem [{\citenamefont {Gross}\ \emph {et~al.}()\citenamefont {Gross}, \citenamefont {Audenaert},\ and\ \citenamefont {Eisert}}]{gross_evenly_2007}%
  \BibitemOpen
  \bibfield  {author} {\bibinfo {author} {\bibfnamefont {D.}~\bibnamefont {Gross}}, \bibinfo {author} {\bibfnamefont {K.}~\bibnamefont {Audenaert}}, \ and\ \bibinfo {author} {\bibfnamefont {J.}~\bibnamefont {Eisert}},\ }\href {\doibase 10.1063/1.2716992} {\ \textbf {\bibinfo {volume} {48}},\ \bibinfo {pages} {052104}},\ \Eprint {http://arxiv.org/abs/quant-ph/0611002} {quant-ph/0611002} \BibitemShut {NoStop}%
\bibitem [{\citenamefont {Pozniak}\ \emph {et~al.}(1998)\citenamefont {Pozniak}, \citenamefont {Zyczkowski},\ and\ \citenamefont {Kus}}]{pozniak1998composed}%
  \BibitemOpen
  \bibfield  {author} {\bibinfo {author} {\bibfnamefont {M.}~\bibnamefont {Pozniak}}, \bibinfo {author} {\bibfnamefont {K.}~\bibnamefont {Zyczkowski}}, \ and\ \bibinfo {author} {\bibfnamefont {M.}~\bibnamefont {Kus}},\ }\href@noop {} {\bibfield  {journal} {\bibinfo  {journal} {Journal of Physics A: Mathematical and General}\ }\textbf {\bibinfo {volume} {31}},\ \bibinfo {pages} {1059} (\bibinfo {year} {1998})}\BibitemShut {NoStop}%
\bibitem [{\citenamefont {Zyczkowski}\ and\ \citenamefont {Kus}()}]{zyczkowski_random_1994}%
  \BibitemOpen
  \bibfield  {author} {\bibinfo {author} {\bibfnamefont {K.}~\bibnamefont {Zyczkowski}}\ and\ \bibinfo {author} {\bibfnamefont {M.}~\bibnamefont {Kus}},\ }\href {\doibase 10.1088/0305-4470/27/12/028} {\ \textbf {\bibinfo {volume} {27}},\ \bibinfo {pages} {4235}}\BibitemShut {NoStop}%
\bibitem [{\citenamefont {Bengtsson}\ and\ \citenamefont {Granstrom}()}]{bengtsson_frame_2008}%
  \BibitemOpen
  \bibfield  {author} {\bibinfo {author} {\bibfnamefont {I.}~\bibnamefont {Bengtsson}}\ and\ \bibinfo {author} {\bibfnamefont {H.}~\bibnamefont {Granstrom}},\ }\href {\doibase 10.48550/arXiv.0808.2947} {\enquote {\bibinfo {title} {The frame potential, on average},}\ }\Eprint {http://arxiv.org/abs/0808.2947 [quant-ph]} {0808.2947 [quant-ph]} \BibitemShut {NoStop}%
\bibitem [{\citenamefont {Brandao}\ \emph {et~al.}(2016)\citenamefont {Brandao}, \citenamefont {Harrow},\ and\ \citenamefont {Horodecki}}]{brandao2016local}%
  \BibitemOpen
  \bibfield  {author} {\bibinfo {author} {\bibfnamefont {F.~G.}\ \bibnamefont {Brandao}}, \bibinfo {author} {\bibfnamefont {A.~W.}\ \bibnamefont {Harrow}}, \ and\ \bibinfo {author} {\bibfnamefont {M.}~\bibnamefont {Horodecki}},\ }\href@noop {} {\bibfield  {journal} {\bibinfo  {journal} {Communications in Mathematical Physics}\ }\textbf {\bibinfo {volume} {346}},\ \bibinfo {pages} {397} (\bibinfo {year} {2016})}\BibitemShut {NoStop}%
\bibitem [{\citenamefont {Harrow}\ and\ \citenamefont {Mehraban}()}]{harrow_approximate_2023}%
  \BibitemOpen
  \bibfield  {author} {\bibinfo {author} {\bibfnamefont {A.}~\bibnamefont {Harrow}}\ and\ \bibinfo {author} {\bibfnamefont {S.}~\bibnamefont {Mehraban}},\ }\href {\doibase 10.1007/s00220-023-04675-z} {\ \textbf {\bibinfo {volume} {401}},\ \bibinfo {pages} {1531}},\ \Eprint {http://arxiv.org/abs/1809.06957 [quant-ph]} {1809.06957 [quant-ph]} \BibitemShut {NoStop}%
\bibitem [{\citenamefont {Du}\ \emph {et~al.}({\natexlab{b}})\citenamefont {Du}, \citenamefont {Hsieh}, \citenamefont {Liu},\ and\ \citenamefont {Tao}}]{du_expressive_2020}%
  \BibitemOpen
  \bibfield  {author} {\bibinfo {author} {\bibfnamefont {Y.}~\bibnamefont {Du}}, \bibinfo {author} {\bibfnamefont {M.-H.}\ \bibnamefont {Hsieh}}, \bibinfo {author} {\bibfnamefont {T.}~\bibnamefont {Liu}}, \ and\ \bibinfo {author} {\bibfnamefont {D.}~\bibnamefont {Tao}},\ }\href {\doibase 10.1103/PhysRevResearch.2.033125} {\ \textbf {\bibinfo {volume} {2}},\ \bibinfo {pages} {033125} ({\natexlab{b}})},\ \Eprint {http://arxiv.org/abs/1810.11922 [quant-ph]} {1810.11922 [quant-ph]} \BibitemShut {NoStop}%
\bibitem [{\citenamefont {Haug}\ \emph {et~al.}()\citenamefont {Haug}, \citenamefont {Bharti},\ and\ \citenamefont {Kim}}]{haug_capacity_2021}%
  \BibitemOpen
  \bibfield  {author} {\bibinfo {author} {\bibfnamefont {T.}~\bibnamefont {Haug}}, \bibinfo {author} {\bibfnamefont {K.}~\bibnamefont {Bharti}}, \ and\ \bibinfo {author} {\bibfnamefont {M.~S.}\ \bibnamefont {Kim}},\ }\href {\doibase 10.1103/PRXQuantum.2.040309} {\ \textbf {\bibinfo {volume} {2}},\ \bibinfo {pages} {040309}},\ \Eprint {http://arxiv.org/abs/2102.01659 [quant-ph, stat]} {2102.01659 [quant-ph, stat]} \BibitemShut {NoStop}%
\bibitem [{\citenamefont {Holmes}\ \emph {et~al.}()\citenamefont {Holmes}, \citenamefont {Sharma}, \citenamefont {Cerezo},\ and\ \citenamefont {Coles}}]{holmes_connecting_2022}%
  \BibitemOpen
  \bibfield  {author} {\bibinfo {author} {\bibfnamefont {Z.}~\bibnamefont {Holmes}}, \bibinfo {author} {\bibfnamefont {K.}~\bibnamefont {Sharma}}, \bibinfo {author} {\bibfnamefont {M.}~\bibnamefont {Cerezo}}, \ and\ \bibinfo {author} {\bibfnamefont {P.~J.}\ \bibnamefont {Coles}},\ }\href {\doibase 10.1103/PRXQuantum.3.010313} {\ \textbf {\bibinfo {volume} {3}},\ \bibinfo {pages} {010313}},\ \Eprint {http://arxiv.org/abs/2101.02138 [quant-ph, stat]} {2101.02138 [quant-ph, stat]} \BibitemShut {NoStop}%
\bibitem [{\citenamefont {Haferkamp}(2022)}]{haferkamp2022random}%
  \BibitemOpen
  \bibfield  {author} {\bibinfo {author} {\bibfnamefont {J.}~\bibnamefont {Haferkamp}},\ }\href@noop {} {\bibfield  {journal} {\bibinfo  {journal} {Quantum}\ }\textbf {\bibinfo {volume} {6}},\ \bibinfo {pages} {795} (\bibinfo {year} {2022})}\BibitemShut {NoStop}%
\bibitem [{\citenamefont {Zhang}\ \emph {et~al.}({\natexlab{a}})\citenamefont {Zhang}, \citenamefont {Bu},\ and\ \citenamefont {Wu}}]{zhang_lower_2016}%
  \BibitemOpen
  \bibfield  {author} {\bibinfo {author} {\bibfnamefont {L.}~\bibnamefont {Zhang}}, \bibinfo {author} {\bibfnamefont {K.}~\bibnamefont {Bu}}, \ and\ \bibinfo {author} {\bibfnamefont {J.}~\bibnamefont {Wu}},\ }\href {\doibase 10.1080/03081087.2015.1057098} {\ \textbf {\bibinfo {volume} {64}},\ \bibinfo {pages} {801} ({\natexlab{a}})},\ \Eprint {http://arxiv.org/abs/1409.3688 [math-ph, physics:quant-ph]} {1409.3688 [math-ph, physics:quant-ph]} \BibitemShut {NoStop}%
\bibitem [{\citenamefont {Dankert}\ \emph {et~al.}(2009)\citenamefont {Dankert}, \citenamefont {Cleve}, \citenamefont {Emerson},\ and\ \citenamefont {Livine}}]{dankert2009exact}%
  \BibitemOpen
  \bibfield  {author} {\bibinfo {author} {\bibfnamefont {C.}~\bibnamefont {Dankert}}, \bibinfo {author} {\bibfnamefont {R.}~\bibnamefont {Cleve}}, \bibinfo {author} {\bibfnamefont {J.}~\bibnamefont {Emerson}}, \ and\ \bibinfo {author} {\bibfnamefont {E.}~\bibnamefont {Livine}},\ }\href@noop {} {\bibfield  {journal} {\bibinfo  {journal} {Physical Review A}\ }\textbf {\bibinfo {volume} {80}},\ \bibinfo {pages} {012304} (\bibinfo {year} {2009})}\BibitemShut {NoStop}%
\bibitem [{\citenamefont {Chen}\ \emph {et~al.}(2021)\citenamefont {Chen}, \citenamefont {Song}, \citenamefont {Zhao},\ and\ \citenamefont {Wang}}]{chen2021variational}%
  \BibitemOpen
  \bibfield  {author} {\bibinfo {author} {\bibfnamefont {R.}~\bibnamefont {Chen}}, \bibinfo {author} {\bibfnamefont {Z.}~\bibnamefont {Song}}, \bibinfo {author} {\bibfnamefont {X.}~\bibnamefont {Zhao}}, \ and\ \bibinfo {author} {\bibfnamefont {X.}~\bibnamefont {Wang}},\ }\href@noop {} {\bibfield  {journal} {\bibinfo  {journal} {Quantum Science and Technology}\ }\textbf {\bibinfo {volume} {7}},\ \bibinfo {pages} {015019} (\bibinfo {year} {2021})}\BibitemShut {NoStop}%
\bibitem [{\citenamefont {Levy}\ \emph {et~al.}(2021)\citenamefont {Levy}, \citenamefont {Luo},\ and\ \citenamefont {Clark}}]{levy2021classical}%
  \BibitemOpen
  \bibfield  {author} {\bibinfo {author} {\bibfnamefont {R.}~\bibnamefont {Levy}}, \bibinfo {author} {\bibfnamefont {D.}~\bibnamefont {Luo}}, \ and\ \bibinfo {author} {\bibfnamefont {B.~K.}\ \bibnamefont {Clark}},\ }\href@noop {} {\bibfield  {journal} {\bibinfo  {journal} {arXiv preprint arXiv:2110.02965}\ } (\bibinfo {year} {2021})}\BibitemShut {NoStop}%
\bibitem [{\citenamefont {Cotler}\ \emph {et~al.}(2017)\citenamefont {Cotler}, \citenamefont {Hunter-Jones}, \citenamefont {Liu},\ and\ \citenamefont {Yoshida}}]{cotler2017chaos}%
  \BibitemOpen
  \bibfield  {author} {\bibinfo {author} {\bibfnamefont {J.}~\bibnamefont {Cotler}}, \bibinfo {author} {\bibfnamefont {N.}~\bibnamefont {Hunter-Jones}}, \bibinfo {author} {\bibfnamefont {J.}~\bibnamefont {Liu}}, \ and\ \bibinfo {author} {\bibfnamefont {B.}~\bibnamefont {Yoshida}},\ }\href@noop {} {\bibfield  {journal} {\bibinfo  {journal} {Journal of High Energy Physics}\ }\textbf {\bibinfo {volume} {2017}},\ \bibinfo {pages} {1} (\bibinfo {year} {2017})}\BibitemShut {NoStop}%
\bibitem [{\citenamefont {{McClean}}\ \emph {et~al.}()\citenamefont {{McClean}}, \citenamefont {Boixo}, \citenamefont {Smelyanskiy}, \citenamefont {Babbush},\ and\ \citenamefont {Neven}}]{mcclean_barren_2018}%
  \BibitemOpen
  \bibfield  {author} {\bibinfo {author} {\bibfnamefont {J.~R.}\ \bibnamefont {{McClean}}}, \bibinfo {author} {\bibfnamefont {S.}~\bibnamefont {Boixo}}, \bibinfo {author} {\bibfnamefont {V.~N.}\ \bibnamefont {Smelyanskiy}}, \bibinfo {author} {\bibfnamefont {R.}~\bibnamefont {Babbush}}, \ and\ \bibinfo {author} {\bibfnamefont {H.}~\bibnamefont {Neven}},\ }\href {\doibase 10.1038/s41467-018-07090-4} {\ \textbf {\bibinfo {volume} {9}},\ \bibinfo {pages} {4812}},\ \Eprint {http://arxiv.org/abs/1803.11173 [physics, physics:quant-ph]} {1803.11173 [physics, physics:quant-ph]} \BibitemShut {NoStop}%
\bibitem [{\citenamefont {Arrasmith}\ \emph {et~al.}()\citenamefont {Arrasmith}, \citenamefont {Holmes}, \citenamefont {Cerezo},\ and\ \citenamefont {Coles}}]{arrasmith_equivalence_2022}%
  \BibitemOpen
  \bibfield  {author} {\bibinfo {author} {\bibfnamefont {A.}~\bibnamefont {Arrasmith}}, \bibinfo {author} {\bibfnamefont {Z.}~\bibnamefont {Holmes}}, \bibinfo {author} {\bibfnamefont {M.}~\bibnamefont {Cerezo}}, \ and\ \bibinfo {author} {\bibfnamefont {P.~J.}\ \bibnamefont {Coles}},\ }\href {\doibase 10.1088/2058-9565/ac7d06} {\ \textbf {\bibinfo {volume} {7}},\ \bibinfo {pages} {045015}},\ \Eprint {http://arxiv.org/abs/2104.05868 [quant-ph]} {2104.05868 [quant-ph]} \BibitemShut {NoStop}%
\bibitem [{\citenamefont {Martín}\ \emph {et~al.}()\citenamefont {Martín}, \citenamefont {Plekhanov},\ and\ \citenamefont {Lubasch}}]{martin_barren_2023}%
  \BibitemOpen
  \bibfield  {author} {\bibinfo {author} {\bibfnamefont {E.}~\bibnamefont {Martín}}, \bibinfo {author} {\bibfnamefont {K.}~\bibnamefont {Plekhanov}}, \ and\ \bibinfo {author} {\bibfnamefont {M.}~\bibnamefont {Lubasch}},\ }\href {\doibase 10.22331/q-2023-04-13-974} {\ \textbf {\bibinfo {volume} {7}},\ \bibinfo {pages} {974}}\BibitemShut {NoStop}%
\bibitem [{\citenamefont {Milsted}\ and\ \citenamefont {Vidal}()}]{milsted_geometric_2018}%
  \BibitemOpen
  \bibfield  {author} {\bibinfo {author} {\bibfnamefont {A.}~\bibnamefont {Milsted}}\ and\ \bibinfo {author} {\bibfnamefont {G.}~\bibnamefont {Vidal}},\ }\href {http://arxiv.org/abs/1812.00529} {\enquote {\bibinfo {title} {Geometric interpretation of the multi-scale entanglement renormalization ansatz},}\ }\Eprint {http://arxiv.org/abs/1812.00529 [cond-mat, physics:hep-th, physics:quant-ph]} {1812.00529 [cond-mat, physics:hep-th, physics:quant-ph]} \BibitemShut {NoStop}%
\bibitem [{\citenamefont {Grant}\ \emph {et~al.}()\citenamefont {Grant}, \citenamefont {Wossnig}, \citenamefont {Ostaszewski},\ and\ \citenamefont {Benedetti}}]{grant_initialization_2019}%
  \BibitemOpen
  \bibfield  {author} {\bibinfo {author} {\bibfnamefont {E.}~\bibnamefont {Grant}}, \bibinfo {author} {\bibfnamefont {L.}~\bibnamefont {Wossnig}}, \bibinfo {author} {\bibfnamefont {M.}~\bibnamefont {Ostaszewski}}, \ and\ \bibinfo {author} {\bibfnamefont {M.}~\bibnamefont {Benedetti}},\ }\href {\doibase 10.22331/q-2019-12-09-214} {\ \textbf {\bibinfo {volume} {3}},\ \bibinfo {pages} {214}},\ \Eprint {http://arxiv.org/abs/1903.05076 [quant-ph]} {1903.05076 [quant-ph]} \BibitemShut {NoStop}%
\bibitem [{\citenamefont {Anand}\ \emph {et~al.}()\citenamefont {Anand}, \citenamefont {Schleich}, \citenamefont {Alperin-Lea}, \citenamefont {Jensen}, \citenamefont {Sim}, \citenamefont {Díaz-Tinoco}, \citenamefont {Kottmann}, \citenamefont {Degroote}, \citenamefont {Izmaylov},\ and\ \citenamefont {Aspuru-Guzik}}]{anand_quantum_2022}%
  \BibitemOpen
  \bibfield  {author} {\bibinfo {author} {\bibfnamefont {A.}~\bibnamefont {Anand}}, \bibinfo {author} {\bibfnamefont {P.}~\bibnamefont {Schleich}}, \bibinfo {author} {\bibfnamefont {S.}~\bibnamefont {Alperin-Lea}}, \bibinfo {author} {\bibfnamefont {P.~W.~K.}\ \bibnamefont {Jensen}}, \bibinfo {author} {\bibfnamefont {S.}~\bibnamefont {Sim}}, \bibinfo {author} {\bibfnamefont {M.}~\bibnamefont {Díaz-Tinoco}}, \bibinfo {author} {\bibfnamefont {J.~S.}\ \bibnamefont {Kottmann}}, \bibinfo {author} {\bibfnamefont {M.}~\bibnamefont {Degroote}}, \bibinfo {author} {\bibfnamefont {A.~F.}\ \bibnamefont {Izmaylov}}, \ and\ \bibinfo {author} {\bibfnamefont {A.}~\bibnamefont {Aspuru-Guzik}},\ }\href {\doibase 10.1039/D1CS00932J} {\ \textbf {\bibinfo {volume} {51}},\ \bibinfo {pages} {1659}},\ \Eprint {http://arxiv.org/abs/2109.15176 [physics, physics:quant-ph]} {2109.15176 [physics, physics:quant-ph]} \BibitemShut {NoStop}%
\bibitem [{\citenamefont {Kremenetski}\ \emph {et~al.}(2021)\citenamefont {Kremenetski}, \citenamefont {Hogg}, \citenamefont {Hadfield}, \citenamefont {Cotton},\ and\ \citenamefont {Tubman}}]{kremenetski2021quantum}%
  \BibitemOpen
  \bibfield  {author} {\bibinfo {author} {\bibfnamefont {V.}~\bibnamefont {Kremenetski}}, \bibinfo {author} {\bibfnamefont {T.}~\bibnamefont {Hogg}}, \bibinfo {author} {\bibfnamefont {S.}~\bibnamefont {Hadfield}}, \bibinfo {author} {\bibfnamefont {S.~J.}\ \bibnamefont {Cotton}}, \ and\ \bibinfo {author} {\bibfnamefont {N.~M.}\ \bibnamefont {Tubman}},\ }\href@noop {} {\bibfield  {journal} {\bibinfo  {journal} {arXiv preprint arXiv:2108.13056}\ } (\bibinfo {year} {2021})}\BibitemShut {NoStop}%
\bibitem [{\citenamefont {Hadfield}\ \emph {et~al.}(2019)\citenamefont {Hadfield}, \citenamefont {Wang}, \citenamefont {O’gorman}, \citenamefont {Rieffel}, \citenamefont {Venturelli},\ and\ \citenamefont {Biswas}}]{hadfield2019quantum}%
  \BibitemOpen
  \bibfield  {author} {\bibinfo {author} {\bibfnamefont {S.}~\bibnamefont {Hadfield}}, \bibinfo {author} {\bibfnamefont {Z.}~\bibnamefont {Wang}}, \bibinfo {author} {\bibfnamefont {B.}~\bibnamefont {O’gorman}}, \bibinfo {author} {\bibfnamefont {E.~G.}\ \bibnamefont {Rieffel}}, \bibinfo {author} {\bibfnamefont {D.}~\bibnamefont {Venturelli}}, \ and\ \bibinfo {author} {\bibfnamefont {R.}~\bibnamefont {Biswas}},\ }\href@noop {} {\bibfield  {journal} {\bibinfo  {journal} {Algorithms}\ }\textbf {\bibinfo {volume} {12}},\ \bibinfo {pages} {34} (\bibinfo {year} {2019})}\BibitemShut {NoStop}%
\bibitem [{\citenamefont {Wiersema}\ \emph {et~al.}({\natexlab{b}})\citenamefont {Wiersema}, \citenamefont {Zhou}, \citenamefont {de~Sereville}, \citenamefont {Carrasquilla}, \citenamefont {Kim},\ and\ \citenamefont {Yuen}}]{wiersema_exploring_2020}%
  \BibitemOpen
  \bibfield  {author} {\bibinfo {author} {\bibfnamefont {R.}~\bibnamefont {Wiersema}}, \bibinfo {author} {\bibfnamefont {C.}~\bibnamefont {Zhou}}, \bibinfo {author} {\bibfnamefont {Y.}~\bibnamefont {de~Sereville}}, \bibinfo {author} {\bibfnamefont {J.~F.}\ \bibnamefont {Carrasquilla}}, \bibinfo {author} {\bibfnamefont {Y.~B.}\ \bibnamefont {Kim}}, \ and\ \bibinfo {author} {\bibfnamefont {H.}~\bibnamefont {Yuen}},\ }\href {\doibase 10.1103/PRXQuantum.1.020319} {\ \textbf {\bibinfo {volume} {1}},\ \bibinfo {pages} {020319} ({\natexlab{b}})},\ \Eprint {http://arxiv.org/abs/2008.02941 [cond-mat, physics:quant-ph]} {2008.02941 [cond-mat, physics:quant-ph]} \BibitemShut {NoStop}%
\bibitem [{\citenamefont {Dborin}\ \emph {et~al.}()\citenamefont {Dborin}, \citenamefont {Barratt}, \citenamefont {Wimalaweera}, \citenamefont {Wright},\ and\ \citenamefont {Green}}]{dborin_matrix_2022}%
  \BibitemOpen
  \bibfield  {author} {\bibinfo {author} {\bibfnamefont {J.}~\bibnamefont {Dborin}}, \bibinfo {author} {\bibfnamefont {F.}~\bibnamefont {Barratt}}, \bibinfo {author} {\bibfnamefont {V.}~\bibnamefont {Wimalaweera}}, \bibinfo {author} {\bibfnamefont {L.}~\bibnamefont {Wright}}, \ and\ \bibinfo {author} {\bibfnamefont {A.~G.}\ \bibnamefont {Green}},\ }\href {\doibase 10.1088/2058-9565/ac7073} {\ \textbf {\bibinfo {volume} {7}},\ \bibinfo {pages} {035014}},\ \bibinfo {note} {publisher: {IOP} Publishing}\BibitemShut {NoStop}%
\bibitem [{noa({\natexlab{b}})}]{noauthor_tensor_nodate}%
  \BibitemOpen
  \href {\doibase 10.1098/rspa.2023.0218} {\enquote {\bibinfo {title} {Tensor networks for quantum machine learning},}\ } ({\natexlab{b}})\BibitemShut {NoStop}%
\bibitem [{\citenamefont {Liu}\ \emph {et~al.}({\natexlab{a}})\citenamefont {Liu}, \citenamefont {Yu}, \citenamefont {Duan},\ and\ \citenamefont {Deng}}]{liu_presence_2022}%
  \BibitemOpen
  \bibfield  {author} {\bibinfo {author} {\bibfnamefont {Z.}~\bibnamefont {Liu}}, \bibinfo {author} {\bibfnamefont {L.-W.}\ \bibnamefont {Yu}}, \bibinfo {author} {\bibfnamefont {L.-M.}\ \bibnamefont {Duan}}, \ and\ \bibinfo {author} {\bibfnamefont {D.-L.}\ \bibnamefont {Deng}},\ }\href {\doibase 10.1103/PhysRevLett.129.270501} {\ \textbf {\bibinfo {volume} {129}},\ \bibinfo {pages} {270501} ({\natexlab{a}})},\ \Eprint {http://arxiv.org/abs/2108.08312 [cond-mat, physics:quant-ph]} {2108.08312 [cond-mat, physics:quant-ph]} \BibitemShut {NoStop}%
\bibitem [{\citenamefont {Martin}\ \emph {et~al.}()\citenamefont {Martin}, \citenamefont {Ayral}, \citenamefont {Jamet}, \citenamefont {Rančić},\ and\ \citenamefont {Simon}}]{martin_combining_2023}%
  \BibitemOpen
  \bibfield  {author} {\bibinfo {author} {\bibfnamefont {B.~A.}\ \bibnamefont {Martin}}, \bibinfo {author} {\bibfnamefont {T.}~\bibnamefont {Ayral}}, \bibinfo {author} {\bibfnamefont {F.}~\bibnamefont {Jamet}}, \bibinfo {author} {\bibfnamefont {M.~J.}\ \bibnamefont {Rančić}}, \ and\ \bibinfo {author} {\bibfnamefont {P.}~\bibnamefont {Simon}},\ }\href {http://arxiv.org/abs/2305.19231} {\enquote {\bibinfo {title} {Combining matrix product states and noisy quantum computers for quantum simulation},}\ }\Eprint {http://arxiv.org/abs/2305.19231 [cond-mat, physics:quant-ph]} {2305.19231 [cond-mat, physics:quant-ph]} \BibitemShut {NoStop}%
\bibitem [{\citenamefont {Zhao}\ and\ \citenamefont {Gao}()}]{zhao_analyzing_2021}%
  \BibitemOpen
  \bibfield  {author} {\bibinfo {author} {\bibfnamefont {C.}~\bibnamefont {Zhao}}\ and\ \bibinfo {author} {\bibfnamefont {X.-S.}\ \bibnamefont {Gao}},\ }\href {\doibase 10.22331/q-2021-06-04-466} {\ \textbf {\bibinfo {volume} {5}},\ \bibinfo {pages} {466}},\ \Eprint {http://arxiv.org/abs/2102.01828 [quant-ph]} {2102.01828 [quant-ph]} \BibitemShut {NoStop}%
\bibitem [{\citenamefont {Anand}\ \emph {et~al.}(2023)\citenamefont {Anand}, \citenamefont {Hauschild}, \citenamefont {Zhang}, \citenamefont {Potter},\ and\ \citenamefont {Zaletel}}]{anand2023holographic}%
  \BibitemOpen
  \bibfield  {author} {\bibinfo {author} {\bibfnamefont {S.}~\bibnamefont {Anand}}, \bibinfo {author} {\bibfnamefont {J.}~\bibnamefont {Hauschild}}, \bibinfo {author} {\bibfnamefont {Y.}~\bibnamefont {Zhang}}, \bibinfo {author} {\bibfnamefont {A.~C.}\ \bibnamefont {Potter}}, \ and\ \bibinfo {author} {\bibfnamefont {M.~P.}\ \bibnamefont {Zaletel}},\ }\href@noop {} {\bibfield  {journal} {\bibinfo  {journal} {PRX Quantum}\ }\textbf {\bibinfo {volume} {4}},\ \bibinfo {pages} {030334} (\bibinfo {year} {2023})}\BibitemShut {NoStop}%
\bibitem [{\citenamefont {Niu}\ \emph {et~al.}(2022)\citenamefont {Niu}, \citenamefont {Haghshenas}, \citenamefont {Zhang}, \citenamefont {Foss-Feig}, \citenamefont {Chan},\ and\ \citenamefont {Potter}}]{PRXQuantum.3.030317}%
  \BibitemOpen
  \bibfield  {author} {\bibinfo {author} {\bibfnamefont {D.}~\bibnamefont {Niu}}, \bibinfo {author} {\bibfnamefont {R.}~\bibnamefont {Haghshenas}}, \bibinfo {author} {\bibfnamefont {Y.}~\bibnamefont {Zhang}}, \bibinfo {author} {\bibfnamefont {M.}~\bibnamefont {Foss-Feig}}, \bibinfo {author} {\bibfnamefont {G.~K.-L.}\ \bibnamefont {Chan}}, \ and\ \bibinfo {author} {\bibfnamefont {A.~C.}\ \bibnamefont {Potter}},\ }\href {\doibase 10.1103/PRXQuantum.3.030317} {\bibfield  {journal} {\bibinfo  {journal} {PRX Quantum}\ }\textbf {\bibinfo {volume} {3}},\ \bibinfo {pages} {030317} (\bibinfo {year} {2022})}\BibitemShut {NoStop}%
\bibitem [{\citenamefont {Haegeman}\ \emph {et~al.}()\citenamefont {Haegeman}, \citenamefont {Mariën}, \citenamefont {Osborne},\ and\ \citenamefont {Verstraete}}]{haegeman_geometry_2014}%
  \BibitemOpen
  \bibfield  {author} {\bibinfo {author} {\bibfnamefont {J.}~\bibnamefont {Haegeman}}, \bibinfo {author} {\bibfnamefont {M.}~\bibnamefont {Mariën}}, \bibinfo {author} {\bibfnamefont {T.~J.}\ \bibnamefont {Osborne}}, \ and\ \bibinfo {author} {\bibfnamefont {F.}~\bibnamefont {Verstraete}},\ }\href {\doibase 10.1063/1.4862851} {\ \textbf {\bibinfo {volume} {55}},\ \bibinfo {pages} {021902}},\ \Eprint {http://arxiv.org/abs/1210.7710 [math-ph, physics:quant-ph]} {1210.7710 [math-ph, physics:quant-ph]} \BibitemShut {NoStop}%
\bibitem [{\citenamefont {Anschuetz}\ and\ \citenamefont {Kiani}()}]{anschuetz_beyond_2022}%
  \BibitemOpen
  \bibfield  {author} {\bibinfo {author} {\bibfnamefont {E.~R.}\ \bibnamefont {Anschuetz}}\ and\ \bibinfo {author} {\bibfnamefont {B.~T.}\ \bibnamefont {Kiani}},\ }\href {\doibase 10.1038/s41467-022-35364-5} {\ \textbf {\bibinfo {volume} {13}},\ \bibinfo {pages} {7760}},\ \Eprint {http://arxiv.org/abs/2205.05786 [quant-ph]} {2205.05786 [quant-ph]} \BibitemShut {NoStop}%
\bibitem [{\citenamefont {Heydari}()}]{heydari_geometric_2016}%
  \BibitemOpen
  \bibfield  {author} {\bibinfo {author} {\bibfnamefont {H.}~\bibnamefont {Heydari}},\ }\href {http://arxiv.org/abs/1503.00238} {\enquote {\bibinfo {title} {Geometric formulation of quantum mechanics},}\ }\Eprint {http://arxiv.org/abs/1503.00238 [math-ph, physics:quant-ph]} {1503.00238 [math-ph, physics:quant-ph]} \BibitemShut {NoStop}%
\bibitem [{\citenamefont {Wootters}(2001)}]{wootters2001entanglement}%
  \BibitemOpen
  \bibfield  {author} {\bibinfo {author} {\bibfnamefont {W.~K.}\ \bibnamefont {Wootters}},\ }\href@noop {} {\bibfield  {journal} {\bibinfo  {journal} {Quantum Inf. Comput.}\ }\textbf {\bibinfo {volume} {1}},\ \bibinfo {pages} {27} (\bibinfo {year} {2001})}\BibitemShut {NoStop}%
\bibitem [{\citenamefont {Katabarwa}\ \emph {et~al.}()\citenamefont {Katabarwa}, \citenamefont {Sim}, \citenamefont {Koh},\ and\ \citenamefont {Dallaire-Demers}}]{katabarwa_connecting_2022}%
  \BibitemOpen
  \bibfield  {author} {\bibinfo {author} {\bibfnamefont {A.}~\bibnamefont {Katabarwa}}, \bibinfo {author} {\bibfnamefont {S.}~\bibnamefont {Sim}}, \bibinfo {author} {\bibfnamefont {D.~E.}\ \bibnamefont {Koh}}, \ and\ \bibinfo {author} {\bibfnamefont {P.-L.}\ \bibnamefont {Dallaire-Demers}},\ }\href {\doibase 10.22331/q-2022-08-23-782} {\ \textbf {\bibinfo {volume} {6}},\ \bibinfo {pages} {782}},\ \Eprint {http://arxiv.org/abs/2106.02593 [quant-ph]} {2106.02593 [quant-ph]} \BibitemShut {NoStop}%
\bibitem [{\citenamefont {Liu}\ \emph {et~al.}({\natexlab{b}})\citenamefont {Liu}, \citenamefont {Yuan}, \citenamefont {Lu},\ and\ \citenamefont {Wang}}]{liu_quantum_2020}%
  \BibitemOpen
  \bibfield  {author} {\bibinfo {author} {\bibfnamefont {J.}~\bibnamefont {Liu}}, \bibinfo {author} {\bibfnamefont {H.}~\bibnamefont {Yuan}}, \bibinfo {author} {\bibfnamefont {X.-M.}\ \bibnamefont {Lu}}, \ and\ \bibinfo {author} {\bibfnamefont {X.}~\bibnamefont {Wang}},\ }\href {\doibase 10.1088/1751-8121/ab5d4d} {\ \textbf {\bibinfo {volume} {53}},\ \bibinfo {pages} {023001} ({\natexlab{b}})},\ \Eprint {http://arxiv.org/abs/1907.08037 [quant-ph]} {1907.08037 [quant-ph]} \BibitemShut {NoStop}%
\bibitem [{\citenamefont {Yamamoto}()}]{yamamoto_natural_2019}%
  \BibitemOpen
  \bibfield  {author} {\bibinfo {author} {\bibfnamefont {N.}~\bibnamefont {Yamamoto}},\ }\href {http://arxiv.org/abs/1909.05074} {\enquote {\bibinfo {title} {On the natural gradient for variational quantum eigensolver},}\ }\Eprint {http://arxiv.org/abs/1909.05074 [quant-ph]} {1909.05074 [quant-ph]} \BibitemShut {NoStop}%
\bibitem [{\citenamefont {Cerezo}\ \emph {et~al.}(2021)\citenamefont {Cerezo}, \citenamefont {Sone}, \citenamefont {Volkoff}, \citenamefont {Cincio},\ and\ \citenamefont {Coles}}]{cerezo2021cost}%
  \BibitemOpen
  \bibfield  {author} {\bibinfo {author} {\bibfnamefont {M.}~\bibnamefont {Cerezo}}, \bibinfo {author} {\bibfnamefont {A.}~\bibnamefont {Sone}}, \bibinfo {author} {\bibfnamefont {T.}~\bibnamefont {Volkoff}}, \bibinfo {author} {\bibfnamefont {L.}~\bibnamefont {Cincio}}, \ and\ \bibinfo {author} {\bibfnamefont {P.~J.}\ \bibnamefont {Coles}},\ }\href@noop {} {\bibfield  {journal} {\bibinfo  {journal} {Nature communications}\ }\textbf {\bibinfo {volume} {12}},\ \bibinfo {pages} {1791} (\bibinfo {year} {2021})}\BibitemShut {NoStop}%
\bibitem [{\citenamefont {Garcia}\ \emph {et~al.}()\citenamefont {Garcia}, \citenamefont {Zhao}, \citenamefont {Bu},\ and\ \citenamefont {Jaffe}}]{garcia_barren_2023}%
  \BibitemOpen
  \bibfield  {author} {\bibinfo {author} {\bibfnamefont {R.~J.}\ \bibnamefont {Garcia}}, \bibinfo {author} {\bibfnamefont {C.}~\bibnamefont {Zhao}}, \bibinfo {author} {\bibfnamefont {K.}~\bibnamefont {Bu}}, \ and\ \bibinfo {author} {\bibfnamefont {A.}~\bibnamefont {Jaffe}},\ }\href {\doibase 10.1007/JHEP01(2023)090} {\ \textbf {\bibinfo {volume} {2023}},\ \bibinfo {pages} {90}},\ \Eprint {http://arxiv.org/abs/2205.06679 [hep-th, physics:quant-ph]} {2205.06679 [hep-th, physics:quant-ph]} \BibitemShut {NoStop}%
\bibitem [{\citenamefont {Liu}\ \emph {et~al.}({\natexlab{c}})\citenamefont {Liu}, \citenamefont {Sun}, \citenamefont {Wu}, \citenamefont {Han},\ and\ \citenamefont {Guo}}]{liu_mitigating_2023}%
  \BibitemOpen
  \bibfield  {author} {\bibinfo {author} {\bibfnamefont {H.-Y.}\ \bibnamefont {Liu}}, \bibinfo {author} {\bibfnamefont {T.-P.}\ \bibnamefont {Sun}}, \bibinfo {author} {\bibfnamefont {Y.-C.}\ \bibnamefont {Wu}}, \bibinfo {author} {\bibfnamefont {Y.-J.}\ \bibnamefont {Han}}, \ and\ \bibinfo {author} {\bibfnamefont {G.-P.}\ \bibnamefont {Guo}},\ }\href {\doibase 10.1088/1367-2630/acb58e} {\ \textbf {\bibinfo {volume} {25}},\ \bibinfo {pages} {013039} ({\natexlab{c}})},\ \bibinfo {note} {publisher: {IOP} Publishing}\BibitemShut {NoStop}%
\bibitem [{\citenamefont {Wiersema}\ \emph {et~al.}({\natexlab{c}})\citenamefont {Wiersema}, \citenamefont {Zhou}, \citenamefont {Carrasquilla},\ and\ \citenamefont {Kim}}]{wiersema_measurement-induced_2023}%
  \BibitemOpen
  \bibfield  {author} {\bibinfo {author} {\bibfnamefont {R.}~\bibnamefont {Wiersema}}, \bibinfo {author} {\bibfnamefont {C.}~\bibnamefont {Zhou}}, \bibinfo {author} {\bibfnamefont {J.~F.}\ \bibnamefont {Carrasquilla}}, \ and\ \bibinfo {author} {\bibfnamefont {Y.~B.}\ \bibnamefont {Kim}},\ }\href {\doibase 10.21468/SciPostPhys.14.6.147} {\ \textbf {\bibinfo {volume} {14}},\ \bibinfo {pages} {147} ({\natexlab{c}})}\BibitemShut {NoStop}%
\bibitem [{\citenamefont {Hastings}(2006)}]{hastings2006solving}%
  \BibitemOpen
  \bibfield  {author} {\bibinfo {author} {\bibfnamefont {M.~B.}\ \bibnamefont {Hastings}},\ }\href@noop {} {\bibfield  {journal} {\bibinfo  {journal} {Physical review b}\ }\textbf {\bibinfo {volume} {73}},\ \bibinfo {pages} {085115} (\bibinfo {year} {2006})}\BibitemShut {NoStop}%
\bibitem [{\citenamefont {Marrero}\ \emph {et~al.}()\citenamefont {Marrero}, \citenamefont {Kieferová},\ and\ \citenamefont {Wiebe}}]{marrero_entanglement_2021}%
  \BibitemOpen
  \bibfield  {author} {\bibinfo {author} {\bibfnamefont {C.~O.}\ \bibnamefont {Marrero}}, \bibinfo {author} {\bibfnamefont {M.}~\bibnamefont {Kieferová}}, \ and\ \bibinfo {author} {\bibfnamefont {N.}~\bibnamefont {Wiebe}},\ }\href {http://arxiv.org/abs/2010.15968} {\enquote {\bibinfo {title} {Entanglement induced barren plateaus},}\ }\Eprint {http://arxiv.org/abs/2010.15968 [quant-ph]} {2010.15968 [quant-ph]} \BibitemShut {NoStop}%
\bibitem [{\citenamefont {Li}\ \emph {et~al.}()\citenamefont {Li}, \citenamefont {Chen},\ and\ \citenamefont {Fisher}}]{li_quantum_2018}%
  \BibitemOpen
  \bibfield  {author} {\bibinfo {author} {\bibfnamefont {Y.}~\bibnamefont {Li}}, \bibinfo {author} {\bibfnamefont {X.}~\bibnamefont {Chen}}, \ and\ \bibinfo {author} {\bibfnamefont {M.~P.~A.}\ \bibnamefont {Fisher}},\ }\href {\doibase 10.1103/PhysRevB.98.205136} {\ \textbf {\bibinfo {volume} {98}},\ \bibinfo {pages} {205136}},\ \Eprint {http://arxiv.org/abs/1808.06134 [cond-mat, physics:quant-ph]} {1808.06134 [cond-mat, physics:quant-ph]} \BibitemShut {NoStop}%
\bibitem [{\citenamefont {Hastings}(2007)}]{hastings2007area}%
  \BibitemOpen
  \bibfield  {author} {\bibinfo {author} {\bibfnamefont {M.~B.}\ \bibnamefont {Hastings}},\ }\href@noop {} {\bibfield  {journal} {\bibinfo  {journal} {Journal of statistical mechanics: theory and experiment}\ }\textbf {\bibinfo {volume} {2007}},\ \bibinfo {pages} {P08024} (\bibinfo {year} {2007})}\BibitemShut {NoStop}%
\bibitem [{\citenamefont {Srednicki}(1994)}]{srednicki1994chaos}%
  \BibitemOpen
  \bibfield  {author} {\bibinfo {author} {\bibfnamefont {M.}~\bibnamefont {Srednicki}},\ }\href@noop {} {\bibfield  {journal} {\bibinfo  {journal} {Physical review e}\ }\textbf {\bibinfo {volume} {50}},\ \bibinfo {pages} {888} (\bibinfo {year} {1994})}\BibitemShut {NoStop}%
\bibitem [{\citenamefont {Sack}\ \emph {et~al.}()\citenamefont {Sack}, \citenamefont {Medina}, \citenamefont {Michailidis}, \citenamefont {Kueng},\ and\ \citenamefont {Serbyn}}]{sack_avoiding_2022}%
  \BibitemOpen
  \bibfield  {author} {\bibinfo {author} {\bibfnamefont {S.~H.}\ \bibnamefont {Sack}}, \bibinfo {author} {\bibfnamefont {R.~A.}\ \bibnamefont {Medina}}, \bibinfo {author} {\bibfnamefont {A.~A.}\ \bibnamefont {Michailidis}}, \bibinfo {author} {\bibfnamefont {R.}~\bibnamefont {Kueng}}, \ and\ \bibinfo {author} {\bibfnamefont {M.}~\bibnamefont {Serbyn}},\ }\href {\doibase 10.1103/PRXQuantum.3.020365} {\ \textbf {\bibinfo {volume} {3}},\ \bibinfo {pages} {020365}},\ \Eprint {http://arxiv.org/abs/2201.08194 [quant-ph]} {2201.08194 [quant-ph]} \BibitemShut {NoStop}%
\bibitem [{\citenamefont {Skolik}\ \emph {et~al.}()\citenamefont {Skolik}, \citenamefont {{McClean}}, \citenamefont {Mohseni}, \citenamefont {van~der Smagt},\ and\ \citenamefont {Leib}}]{skolik_layerwise_2021}%
  \BibitemOpen
  \bibfield  {author} {\bibinfo {author} {\bibfnamefont {A.}~\bibnamefont {Skolik}}, \bibinfo {author} {\bibfnamefont {J.~R.}\ \bibnamefont {{McClean}}}, \bibinfo {author} {\bibfnamefont {M.}~\bibnamefont {Mohseni}}, \bibinfo {author} {\bibfnamefont {P.}~\bibnamefont {van~der Smagt}}, \ and\ \bibinfo {author} {\bibfnamefont {M.}~\bibnamefont {Leib}},\ }\href {\doibase 10.1007/s42484-020-00036-4} {\ \textbf {\bibinfo {volume} {3}},\ \bibinfo {pages} {5}},\ \Eprint {http://arxiv.org/abs/2006.14904 [quant-ph]} {2006.14904 [quant-ph]} \BibitemShut {NoStop}%
\bibitem [{\citenamefont {Sharma}\ \emph {et~al.}()\citenamefont {Sharma}, \citenamefont {Cerezo}, \citenamefont {Cincio},\ and\ \citenamefont {Coles}}]{sharma_trainability_2022}%
  \BibitemOpen
  \bibfield  {author} {\bibinfo {author} {\bibfnamefont {K.}~\bibnamefont {Sharma}}, \bibinfo {author} {\bibfnamefont {M.}~\bibnamefont {Cerezo}}, \bibinfo {author} {\bibfnamefont {L.}~\bibnamefont {Cincio}}, \ and\ \bibinfo {author} {\bibfnamefont {P.~J.}\ \bibnamefont {Coles}},\ }\href {\doibase 10.1103/PhysRevLett.128.180505} {\ \textbf {\bibinfo {volume} {128}},\ \bibinfo {pages} {180505}},\ \Eprint {http://arxiv.org/abs/2005.12458 [quant-ph]} {2005.12458 [quant-ph]} \BibitemShut {NoStop}%
\bibitem [{\citenamefont {Zhang}\ \emph {et~al.}({\natexlab{b}})\citenamefont {Zhang}, \citenamefont {Bengio}, \citenamefont {Hardt}, \citenamefont {Recht},\ and\ \citenamefont {Vinyals}}]{zhang_understanding_2017}%
  \BibitemOpen
  \bibfield  {author} {\bibinfo {author} {\bibfnamefont {C.}~\bibnamefont {Zhang}}, \bibinfo {author} {\bibfnamefont {S.}~\bibnamefont {Bengio}}, \bibinfo {author} {\bibfnamefont {M.}~\bibnamefont {Hardt}}, \bibinfo {author} {\bibfnamefont {B.}~\bibnamefont {Recht}}, \ and\ \bibinfo {author} {\bibfnamefont {O.}~\bibnamefont {Vinyals}},\ }\href {http://arxiv.org/abs/1611.03530} {\enquote {\bibinfo {title} {Understanding deep learning requires rethinking generalization},}\ } ({\natexlab{b}}),\ \Eprint {http://arxiv.org/abs/1611.03530 [cs]} {1611.03530 [cs]} \BibitemShut {NoStop}%
\bibitem [{\citenamefont {Raghu}\ \emph {et~al.}(2017)\citenamefont {Raghu}, \citenamefont {Poole}, \citenamefont {Kleinberg}, \citenamefont {Ganguli},\ and\ \citenamefont {Sohl-Dickstein}}]{raghu2017expressive}%
  \BibitemOpen
  \bibfield  {author} {\bibinfo {author} {\bibfnamefont {M.}~\bibnamefont {Raghu}}, \bibinfo {author} {\bibfnamefont {B.}~\bibnamefont {Poole}}, \bibinfo {author} {\bibfnamefont {J.}~\bibnamefont {Kleinberg}}, \bibinfo {author} {\bibfnamefont {S.}~\bibnamefont {Ganguli}}, \ and\ \bibinfo {author} {\bibfnamefont {J.}~\bibnamefont {Sohl-Dickstein}},\ }in\ \href@noop {} {\emph {\bibinfo {booktitle} {international conference on machine learning}}}\ (\bibinfo {organization} {PMLR},\ \bibinfo {year} {2017})\ pp.\ \bibinfo {pages} {2847--2854}\BibitemShut {NoStop}%
\bibitem [{\citenamefont {Roberts}\ \emph {et~al.}()\citenamefont {Roberts}, \citenamefont {Yaida},\ and\ \citenamefont {Hanin}}]{roberts_principles_2022}%
  \BibitemOpen
  \bibfield  {author} {\bibinfo {author} {\bibfnamefont {D.~A.}\ \bibnamefont {Roberts}}, \bibinfo {author} {\bibfnamefont {S.}~\bibnamefont {Yaida}}, \ and\ \bibinfo {author} {\bibfnamefont {B.}~\bibnamefont {Hanin}},\ }\href {\doibase 10.1017/9781009023405} {\emph {\bibinfo {title} {The Principles of Deep Learning Theory}}},\ \Eprint {http://arxiv.org/abs/2106.10165 [hep-th, stat]} {2106.10165 [hep-th, stat]} \BibitemShut {NoStop}%
\bibitem [{\citenamefont {Gonon}\ and\ \citenamefont {Jacquier}()}]{gonon_universal_2023}%
  \BibitemOpen
  \bibfield  {author} {\bibinfo {author} {\bibfnamefont {L.}~\bibnamefont {Gonon}}\ and\ \bibinfo {author} {\bibfnamefont {A.}~\bibnamefont {Jacquier}},\ }\href {http://arxiv.org/abs/2307.12904} {\enquote {\bibinfo {title} {Universal approximation theorem and error bounds for quantum neural networks and quantum reservoirs},}\ }\Eprint {http://arxiv.org/abs/2307.12904 [quant-ph]} {2307.12904 [quant-ph]} \BibitemShut {NoStop}%
\bibitem [{\citenamefont {Pesah}\ \emph {et~al.}()\citenamefont {Pesah}, \citenamefont {Cerezo}, \citenamefont {Wang}, \citenamefont {Volkoff}, \citenamefont {Sornborger},\ and\ \citenamefont {Coles}}]{pesah_absence_2021}%
  \BibitemOpen
  \bibfield  {author} {\bibinfo {author} {\bibfnamefont {A.}~\bibnamefont {Pesah}}, \bibinfo {author} {\bibfnamefont {M.}~\bibnamefont {Cerezo}}, \bibinfo {author} {\bibfnamefont {S.}~\bibnamefont {Wang}}, \bibinfo {author} {\bibfnamefont {T.}~\bibnamefont {Volkoff}}, \bibinfo {author} {\bibfnamefont {A.~T.}\ \bibnamefont {Sornborger}}, \ and\ \bibinfo {author} {\bibfnamefont {P.~J.}\ \bibnamefont {Coles}},\ }\href {\doibase 10.1103/PhysRevX.11.041011} {\ \textbf {\bibinfo {volume} {11}},\ \bibinfo {pages} {041011}},\ \Eprint {http://arxiv.org/abs/2011.02966 [quant-ph, stat]} {2011.02966 [quant-ph, stat]} \BibitemShut {NoStop}%
\bibitem [{\citenamefont {Kjaergaard}\ \emph {et~al.}(2020)\citenamefont {Kjaergaard}, \citenamefont {Schwartz}, \citenamefont {Braum{\"u}ller}, \citenamefont {Krantz}, \citenamefont {Wang}, \citenamefont {Gustavsson},\ and\ \citenamefont {Oliver}}]{kjaergaard2020superconducting}%
  \BibitemOpen
  \bibfield  {author} {\bibinfo {author} {\bibfnamefont {M.}~\bibnamefont {Kjaergaard}}, \bibinfo {author} {\bibfnamefont {M.~E.}\ \bibnamefont {Schwartz}}, \bibinfo {author} {\bibfnamefont {J.}~\bibnamefont {Braum{\"u}ller}}, \bibinfo {author} {\bibfnamefont {P.}~\bibnamefont {Krantz}}, \bibinfo {author} {\bibfnamefont {J.~I.-J.}\ \bibnamefont {Wang}}, \bibinfo {author} {\bibfnamefont {S.}~\bibnamefont {Gustavsson}}, \ and\ \bibinfo {author} {\bibfnamefont {W.~D.}\ \bibnamefont {Oliver}},\ }\href@noop {} {\bibfield  {journal} {\bibinfo  {journal} {Annual Review of Condensed Matter Physics}\ }\textbf {\bibinfo {volume} {11}},\ \bibinfo {pages} {369} (\bibinfo {year} {2020})}\BibitemShut {NoStop}%
\bibitem [{\citenamefont {Bruzewicz}\ \emph {et~al.}(2019)\citenamefont {Bruzewicz}, \citenamefont {Chiaverini}, \citenamefont {McConnell},\ and\ \citenamefont {Sage}}]{bruzewicz2019trapped}%
  \BibitemOpen
  \bibfield  {author} {\bibinfo {author} {\bibfnamefont {C.~D.}\ \bibnamefont {Bruzewicz}}, \bibinfo {author} {\bibfnamefont {J.}~\bibnamefont {Chiaverini}}, \bibinfo {author} {\bibfnamefont {R.}~\bibnamefont {McConnell}}, \ and\ \bibinfo {author} {\bibfnamefont {J.~M.}\ \bibnamefont {Sage}},\ }\href@noop {} {\bibfield  {journal} {\bibinfo  {journal} {Applied Physics Reviews}\ }\textbf {\bibinfo {volume} {6}} (\bibinfo {year} {2019})}\BibitemShut {NoStop}%
\bibitem [{\citenamefont {Araz}\ and\ \citenamefont {Spannowsky}(2022)}]{araz2022classical}%
  \BibitemOpen
  \bibfield  {author} {\bibinfo {author} {\bibfnamefont {J.~Y.}\ \bibnamefont {Araz}}\ and\ \bibinfo {author} {\bibfnamefont {M.}~\bibnamefont {Spannowsky}},\ }\href@noop {} {\bibfield  {journal} {\bibinfo  {journal} {Physical Review A}\ }\textbf {\bibinfo {volume} {106}},\ \bibinfo {pages} {062423} (\bibinfo {year} {2022})}\BibitemShut {NoStop}%
\bibitem [{\citenamefont {Tan}\ and\ \citenamefont {Lim}(2019)}]{tan2019vanishing}%
  \BibitemOpen
  \bibfield  {author} {\bibinfo {author} {\bibfnamefont {H.~H.}\ \bibnamefont {Tan}}\ and\ \bibinfo {author} {\bibfnamefont {K.~H.}\ \bibnamefont {Lim}},\ }in\ \href@noop {} {\emph {\bibinfo {booktitle} {2019 7th international conference on smart computing \& communications (ICSCC)}}}\ (\bibinfo {organization} {IEEE},\ \bibinfo {year} {2019})\ pp.\ \bibinfo {pages} {1--4}\BibitemShut {NoStop}%
\bibitem [{\citenamefont {Larocca}\ \emph {et~al.}({\natexlab{a}})\citenamefont {Larocca}, \citenamefont {Czarnik}, \citenamefont {Sharma}, \citenamefont {Muraleedharan}, \citenamefont {Coles},\ and\ \citenamefont {Cerezo}}]{larocca_diagnosing_2022}%
  \BibitemOpen
  \bibfield  {author} {\bibinfo {author} {\bibfnamefont {M.}~\bibnamefont {Larocca}}, \bibinfo {author} {\bibfnamefont {P.}~\bibnamefont {Czarnik}}, \bibinfo {author} {\bibfnamefont {K.}~\bibnamefont {Sharma}}, \bibinfo {author} {\bibfnamefont {G.}~\bibnamefont {Muraleedharan}}, \bibinfo {author} {\bibfnamefont {P.~J.}\ \bibnamefont {Coles}}, \ and\ \bibinfo {author} {\bibfnamefont {M.}~\bibnamefont {Cerezo}},\ }\href {\doibase 10.22331/q-2022-09-29-824} {\ \textbf {\bibinfo {volume} {6}},\ \bibinfo {pages} {824} ({\natexlab{a}})},\ \bibinfo {note} {publisher: Verein zur Förderung des Open Access Publizierens in den Quantenwissenschaften}\BibitemShut {NoStop}%
\bibitem [{\citenamefont {LeCun}\ \emph {et~al.}(1988)\citenamefont {LeCun}, \citenamefont {Touresky}, \citenamefont {Hinton},\ and\ \citenamefont {Sejnowski}}]{lecun1988theoretical}%
  \BibitemOpen
  \bibfield  {author} {\bibinfo {author} {\bibfnamefont {Y.}~\bibnamefont {LeCun}}, \bibinfo {author} {\bibfnamefont {D.}~\bibnamefont {Touresky}}, \bibinfo {author} {\bibfnamefont {G.}~\bibnamefont {Hinton}}, \ and\ \bibinfo {author} {\bibfnamefont {T.}~\bibnamefont {Sejnowski}},\ }in\ \href@noop {} {\emph {\bibinfo {booktitle} {Proceedings of the 1988 connectionist models summer school}}},\ Vol.~\bibinfo {volume} {1}\ (\bibinfo {organization} {San Mateo, CA, USA},\ \bibinfo {year} {1988})\ pp.\ \bibinfo {pages} {21--28}\BibitemShut {NoStop}%
\bibitem [{\citenamefont {Hanin}(2018)}]{hanin2018neural}%
  \BibitemOpen
  \bibfield  {author} {\bibinfo {author} {\bibfnamefont {B.}~\bibnamefont {Hanin}},\ }\href@noop {} {\bibfield  {journal} {\bibinfo  {journal} {Advances in neural information processing systems}\ }\textbf {\bibinfo {volume} {31}} (\bibinfo {year} {2018})}\BibitemShut {NoStop}%
\bibitem [{\citenamefont {Pucha{\l}a}\ and\ \citenamefont {Miszczak}(2011)}]{puchala2011symbolic}%
  \BibitemOpen
  \bibfield  {author} {\bibinfo {author} {\bibfnamefont {Z.}~\bibnamefont {Pucha{\l}a}}\ and\ \bibinfo {author} {\bibfnamefont {J.~A.}\ \bibnamefont {Miszczak}},\ }\href@noop {} {\bibfield  {journal} {\bibinfo  {journal} {arXiv preprint arXiv:1109.4244}\ } (\bibinfo {year} {2011})}\BibitemShut {NoStop}%
\bibitem [{\citenamefont {Arora}\ \emph {et~al.}()\citenamefont {Arora}, \citenamefont {Du}, \citenamefont {Hu}, \citenamefont {Li}, \citenamefont {Salakhutdinov},\ and\ \citenamefont {Wang}}]{arora_exact_2019}%
  \BibitemOpen
  \bibfield  {author} {\bibinfo {author} {\bibfnamefont {S.}~\bibnamefont {Arora}}, \bibinfo {author} {\bibfnamefont {S.~S.}\ \bibnamefont {Du}}, \bibinfo {author} {\bibfnamefont {W.}~\bibnamefont {Hu}}, \bibinfo {author} {\bibfnamefont {Z.}~\bibnamefont {Li}}, \bibinfo {author} {\bibfnamefont {R.}~\bibnamefont {Salakhutdinov}}, \ and\ \bibinfo {author} {\bibfnamefont {R.}~\bibnamefont {Wang}},\ }\href {http://arxiv.org/abs/1904.11955} {\enquote {\bibinfo {title} {On exact computation with an infinitely wide neural net},}\ }\Eprint {http://arxiv.org/abs/1904.11955 [cs, stat]} {1904.11955 [cs, stat]} \BibitemShut {NoStop}%
\bibitem [{\citenamefont {Larocca}\ \emph {et~al.}({\natexlab{b}})\citenamefont {Larocca}, \citenamefont {Ju}, \citenamefont {García-Martín}, \citenamefont {Coles},\ and\ \citenamefont {Cerezo}}]{larocca_theory_2023}%
  \BibitemOpen
  \bibfield  {author} {\bibinfo {author} {\bibfnamefont {M.}~\bibnamefont {Larocca}}, \bibinfo {author} {\bibfnamefont {N.}~\bibnamefont {Ju}}, \bibinfo {author} {\bibfnamefont {D.}~\bibnamefont {García-Martín}}, \bibinfo {author} {\bibfnamefont {P.~J.}\ \bibnamefont {Coles}}, \ and\ \bibinfo {author} {\bibfnamefont {M.}~\bibnamefont {Cerezo}},\ }\href {\doibase 10.1038/s43588-023-00467-6} {\ \textbf {\bibinfo {volume} {3}},\ \bibinfo {pages} {542} ({\natexlab{b}})},\ \bibinfo {note} {number: 6 Publisher: Nature Publishing Group}\BibitemShut {NoStop}%
\bibitem [{\citenamefont {Abedi}\ \emph {et~al.}()\citenamefont {Abedi}, \citenamefont {Beigi},\ and\ \citenamefont {Taghavi}}]{abedi_quantum_2023}%
  \BibitemOpen
  \bibfield  {author} {\bibinfo {author} {\bibfnamefont {E.}~\bibnamefont {Abedi}}, \bibinfo {author} {\bibfnamefont {S.}~\bibnamefont {Beigi}}, \ and\ \bibinfo {author} {\bibfnamefont {L.}~\bibnamefont {Taghavi}},\ }\href {\doibase 10.22331/q-2023-04-27-989} {\ \textbf {\bibinfo {volume} {7}},\ \bibinfo {pages} {989}},\ \Eprint {http://arxiv.org/abs/2202.08232 [quant-ph]} {2202.08232 [quant-ph]} \BibitemShut {NoStop}%
\bibitem [{\citenamefont {Jacot}\ \emph {et~al.}(2018)\citenamefont {Jacot}, \citenamefont {Gabriel},\ and\ \citenamefont {Hongler}}]{jacot2018neural}%
  \BibitemOpen
  \bibfield  {author} {\bibinfo {author} {\bibfnamefont {A.}~\bibnamefont {Jacot}}, \bibinfo {author} {\bibfnamefont {F.}~\bibnamefont {Gabriel}}, \ and\ \bibinfo {author} {\bibfnamefont {C.}~\bibnamefont {Hongler}},\ }\href@noop {} {\bibfield  {journal} {\bibinfo  {journal} {Advances in neural information processing systems}\ }\textbf {\bibinfo {volume} {31}} (\bibinfo {year} {2018})}\BibitemShut {NoStop}%
\bibitem [{\citenamefont {Liu}\ \emph {et~al.}(2020)\citenamefont {Liu}, \citenamefont {Zhu},\ and\ \citenamefont {Belkin}}]{liu2020linearity}%
  \BibitemOpen
  \bibfield  {author} {\bibinfo {author} {\bibfnamefont {C.}~\bibnamefont {Liu}}, \bibinfo {author} {\bibfnamefont {L.}~\bibnamefont {Zhu}}, \ and\ \bibinfo {author} {\bibfnamefont {M.}~\bibnamefont {Belkin}},\ }\href@noop {} {\bibfield  {journal} {\bibinfo  {journal} {Advances in Neural Information Processing Systems}\ }\textbf {\bibinfo {volume} {33}},\ \bibinfo {pages} {15954} (\bibinfo {year} {2020})}\BibitemShut {NoStop}%
\bibitem [{\citenamefont {Liu}\ \emph {et~al.}({\natexlab{d}})\citenamefont {Liu}, \citenamefont {Tacchino}, \citenamefont {Glick}, \citenamefont {Jiang},\ and\ \citenamefont {Mezzacapo}}]{liu_representation_2022}%
  \BibitemOpen
  \bibfield  {author} {\bibinfo {author} {\bibfnamefont {J.}~\bibnamefont {Liu}}, \bibinfo {author} {\bibfnamefont {F.}~\bibnamefont {Tacchino}}, \bibinfo {author} {\bibfnamefont {J.~R.}\ \bibnamefont {Glick}}, \bibinfo {author} {\bibfnamefont {L.}~\bibnamefont {Jiang}}, \ and\ \bibinfo {author} {\bibfnamefont {A.}~\bibnamefont {Mezzacapo}},\ }\href {\doibase 10.1103/PRXQuantum.3.030323} {\ \textbf {\bibinfo {volume} {3}},\ \bibinfo {pages} {030323} ({\natexlab{d}})},\ \Eprint {http://arxiv.org/abs/2111.04225 [quant-ph, stat]} {2111.04225 [quant-ph, stat]} \BibitemShut {NoStop}%
\bibitem [{\citenamefont {Nakaji}\ \emph {et~al.}()\citenamefont {Nakaji}, \citenamefont {Tezuka},\ and\ \citenamefont {Yamamoto}}]{nakaji_quantum-enhanced_2023}%
  \BibitemOpen
  \bibfield  {author} {\bibinfo {author} {\bibfnamefont {K.}~\bibnamefont {Nakaji}}, \bibinfo {author} {\bibfnamefont {H.}~\bibnamefont {Tezuka}}, \ and\ \bibinfo {author} {\bibfnamefont {N.}~\bibnamefont {Yamamoto}},\ }\href {http://arxiv.org/abs/2109.03786} {\enquote {\bibinfo {title} {Quantum-enhanced neural networks in the neural tangent kernel framework},}\ }\Eprint {http://arxiv.org/abs/2109.03786 [quant-ph]} {2109.03786 [quant-ph]} \BibitemShut {NoStop}%
\bibitem [{\citenamefont {Liu}\ \emph {et~al.}({\natexlab{e}})\citenamefont {Liu}, \citenamefont {Lin},\ and\ \citenamefont {Jiang}}]{liu_laziness_2022}%
  \BibitemOpen
  \bibfield  {author} {\bibinfo {author} {\bibfnamefont {J.}~\bibnamefont {Liu}}, \bibinfo {author} {\bibfnamefont {Z.}~\bibnamefont {Lin}}, \ and\ \bibinfo {author} {\bibfnamefont {L.}~\bibnamefont {Jiang}},\ }\href {http://arxiv.org/abs/2206.09313} {\enquote {\bibinfo {title} {Laziness, barren plateau, and noise in machine learning},}\ } ({\natexlab{e}}),\ \Eprint {http://arxiv.org/abs/2206.09313 [quant-ph, stat]} {2206.09313 [quant-ph, stat]} \BibitemShut {NoStop}%
\bibitem [{\citenamefont {Bouland}\ \emph {et~al.}(2019)\citenamefont {Bouland}, \citenamefont {Fefferman}, \citenamefont {Nirkhe},\ and\ \citenamefont {Vazirani}}]{bouland2019complexity}%
  \BibitemOpen
  \bibfield  {author} {\bibinfo {author} {\bibfnamefont {A.}~\bibnamefont {Bouland}}, \bibinfo {author} {\bibfnamefont {B.}~\bibnamefont {Fefferman}}, \bibinfo {author} {\bibfnamefont {C.}~\bibnamefont {Nirkhe}}, \ and\ \bibinfo {author} {\bibfnamefont {U.}~\bibnamefont {Vazirani}},\ }\href@noop {} {\bibfield  {journal} {\bibinfo  {journal} {Nature Physics}\ }\textbf {\bibinfo {volume} {15}},\ \bibinfo {pages} {159} (\bibinfo {year} {2019})}\BibitemShut {NoStop}%
\end{thebibliography}%
\end{document}